\documentclass[useAMS,usenatbib]{mn2e}
\usepackage{epsfig}
 at 10truept

\title
     [Radiative feedback from an early X-ray background]
{\vglue-3.0truecm
\vglue 2.5truecm
  Radiative feedback from an early X-ray background
\author
  [S.C.O. Glover \& P.W.J.L. Brand]
  { S.C.O. Glover,$^{\! 1,2}$ \& P.W.J.L. Brand$^{\! 1}$\\
  \vspace*{1mm}\\
  $^1$ Institute for Astronomy, 
       University of Edinburgh,
       Royal Observatory,
       Blackford Hill, 
       Edinburgh, 
       EH9 3HJ \\
  $^2$ Department of Astrophysics, 
       American Museum of Natural History, 
       Central Park West at 79th Street, 
       New York, 
       NY 10024
  } 
}

\ifCUPmtlplainloaded \else
 
\fi

\newcommand{\Hm}{\rm{H}^{-}}
\newcommand{\Hp}{\rm{H}^{+}}
\newcommand{\Hep}{\rm{He}^{+}}
\newcommand{\Hepp}{\rm{He}^{++}}
\newcommand{\me}{\rm{e}}
\newcommand{\mH}{\rm{H}}
\newcommand{\He}{\rm{He}}
\newcommand{\mHt}{\rm{H}_{2}}
\newcommand{\mHtp}{\rm{H}_{2}^{+}}
\newcommand{\deriv}[1]{\,{\rm d}#1}
\newcommand{\hi}{\hbox{H\,{\sc i}}\,}                    
\newcommand{\hii}{\hbox{H\,{\sc ii}}\,}
\newcommand{\hei}{\hbox{He\,{\sc i}}\,}
\newcommand{\heii}{\hbox{He\,{\sc ii}}\,}
\newcommand{\heiii}{\hbox{He\,{\sc iii}}\,}
\def\simless{\mathbin{\lower 3pt\hbox
   {$\rlap{\raise 5pt\hbox{$\char'074$}}\mathchar"7218$}}}   
\def\simgreat{\mathbin{\lower 3pt\hbox  
   {$\rlap{\raise 5pt\hbox{$\char'076$}}\mathchar"7218$}}}   

\begin{document}

\maketitle

\begin{abstract}
The first generation of stars (commonly known as population III) are 
expected to form in low-mass protogalaxies in which molecular hydrogen
is the dominant coolant. Radiation from these stars will rapidly build
up an extragalactic ultraviolet background capable of photodissociating 
$\mHt$, and it is widely believed that this background will suppress
further star formation in low-mass systems.

However, star formation will also produce an extragalactic X-ray 
background. This X-ray background, by increasing the fractional 
ionization of protogalactic gas, promotes $\mHt$ formation and 
reduces the effectiveness of ultraviolet feedback.

In this paper, we examine which of these backgrounds has the dominant
effect. Using a simple model for the growth of the UV and X-ray backgrounds,
together with a detailed one-dimensional model of protogalactic chemical
evolution, we examine the effects of the X-ray backgrounds produced by a 
number of likely source models. We show that in several cases,
the resulting X-ray background is strong enough to offset UV
photodissociation in large $\mHt$-cooled protogalaxies. On the other
hand, small protogalaxies (those with virial temperatures 
$T_{\rm vir} < 2000 \: \rm{K}$) remain dominated by the UV background in all
of the models we examine.

We also briefly investigate the effects of the X-ray background
upon the thermal and chemical evolution of the diffuse IGM.
\end{abstract}


\section{Introduction}
In cosmological models based on cold dark matter (CDM),
the first stars are believed to form within small protogalaxies, with 
virial temperatures $T_{\rm vir} < 10^{4} \: \rm{K}$ \citep{cr,htl,teg}.
Cooling within these protogalaxies is dominated by molecular hydrogen, 
$\mHt$, which forms via the gas-phase reactions 
\begin{eqnarray}
 \mH + \me & \rightarrow & \Hm + \gamma  \label{h2f1} \\
 \Hm + \mH & \rightarrow & \mHt + \me, 
\end{eqnarray}
and
\begin{eqnarray}
 \mH + \Hp & \rightarrow & \mHtp + \gamma \\
 \mHtp + \mH & \rightarrow & \mHt + \Hp, \label{h2f4}
\end{eqnarray}
even in the absence of dust. Although the fractional abundance of 
$\mHt$ that forms in this way is small, it is sufficient to allow for 
effective cooling and the formation of stars \citep{htl,teg,bcl,abn}.

As soon as massive stars form, however, they immediately begin to 
photoionize and photodissociate this $\mHt$. Photoionization requires 
photons with energies greater than $15.4 \: \rm{eV}$, which are 
strongly absorbed by neutral hydrogen, and is only of importance within
\hii regions. Photodissociation, by contrast, occurs through the 
absorption of photons in the Lyman-Werner band system \citep{sw}, 
with energies in the range
$11.18$ -- $13.6 \: \rm{eV}$. These photons are not strongly absorbed by
neutral hydrogen and can readily escape into the intergalactic medium 
\citep{on,scog}. 

Initially, many of these photons will be absorbed by intergalactic $\mHt$,
but its abundance is small and it is rapidly photodissociated 
(see section~\ref{ther_chem} below). Consequently, the onset of star 
formation is 
soon followed by the appearance of an ultraviolet background radiation 
field. This ultraviolet background acts to suppress further star formation by 
photodissociating $\mHt$ within newly-forming protogalaxies. 
The effects of this background have been studied by a number of authors
\citep*{cfgj,cfa,hrlx,hrl97,har,mba}. In particular, \citet{har} 
study the coupled problem of the evolution of the ultraviolet background 
and its feedback on the global star formation rate using a simple 
galaxy formation model based on the Press-Schechter formalism \citep{ps}. 
They find that cooling (and hence star formation) within small protogalaxies
is completely suppressed prior to cosmological reionization. 
Taken at face value, their results suggest that star formation 
within small, $\mHt$-cooled protogalaxies is a transient phenomenon, 
with little impact on later stages of galaxy formation.

These conclusions, however, rest on the assumption that the only free 
electrons present in the protogalactic gas come from the small residual 
fraction remaining after cosmological recombination. This is important,
as free electrons (and protons) catalyze $\mHt$ formation, as we can see
from equations~\ref{h2f1} to \ref{h2f4}. If the free electron abundance were
significantly higher, then the $\mHt$ formation rate would also be higher, 
offsetting the effects of photodissociation. At the very least,  this 
would delay the suppression of star formation, and in principle could 
entirely negate it.

It is therefore important to determine whether there is any way in which
an enhanced level of ionization could be produced. 
One possible source is the residual ionization that would remain after the
recombination of \hii regions produced by an earlier generation of stars. 
This has recently been studied by 
\citet{rgs,rgs2}, who find that it can be an effective source of $\mHt$
and can dramatically reduce the effectiveness of photodissociation feedback.
However, their conclusions are still somewhat uncertain, both because their
simulations are under-resolved \citep[see figures 5--7 in][]{rgs} and because
they neglect the effects of supernovae, which would act to disperse the gas 
and significantly lengthen the recombination timescale, thereby delaying
the formation of $\mHt$. 

An alternative possibility is ionization by a high redshift X-ray 
background. At X-ray energies, the optical depth of the intergalactic medium
(IGM) is small and any X-ray sources present will naturally generate an 
X-ray background. Moreover, X-rays can penetrate to large depths within 
newly-formed protogalaxies, allowing them to raise the fractional ionization 
throughout the gas.

The potential importance of such a background was first highlighted 
by \citet{har}, with high redshift quasars suggested as a possible source.
Using a very simple model for a quasar-produced X-ray background, they showed 
that if quasars contribute more than 10\% of the UV background,
then the ionization produced by the associated X-ray background is sufficient
to negate the effects of UV photodissociation. Indeed, they found evidence that
such a background could actually promote cooling within the dense gas in the 
centres of protogalaxies. This scenario has not yet been firmly ruled out, but
observational evidence suggests that quasars are unlikely to be present in
sufficient number at high redshift \citep{pei,hml}. 

Quasars, however, are not the only potential source of X-rays. 
Star formation also leads to the production of X-rays,
 primarily through the formation of massive X-ray binaries \citep{helf}, but 
also through bremsstrahlung and inverse Compton emission from supernova 
remnants. These sources generate only a small fraction of the present-day 
X-ray background \citep{an}, but may become dominant at high redshifts. 

In this paper, we examine the effects of the X-rays produced by these sources,
both on the cooling of gas within virialized protogalaxies and also on
the chemical and thermal evolution of the IGM. The outline of the
paper is as follows. In section~\ref{back}, we discuss the sources responsible
for producing the UV and X-ray backgrounds, and show how the build-up of these
backgrounds can be computed. In section~\ref{meth}, we outline the method
used to study the effect of this radiation on the primordial gas, 
and in section~\ref{res} apply it for a number of different X-ray source
models. We present our conclusions in section~\ref{conc}.

\section{The UV and X-ray backgrounds}
\label{back}
For an observer at redshift $z_{0}$, we can write the mean specific 
intensity of the radiation background at an observed frequency $\nu_{0}$
as \citep{mhr}
\begin{equation}
\label{bg}
 J(\nu_{0},z_{0}) = \frac{1}{4\pi} \int^{\infty}_{z_{0}} 
\varepsilon(\nu,z) e^{-\tau(\nu_{0},z_{0},z)}
\frac{(1+z_{0})^{3}}{(1+z)^{3}} 
\frac{\deriv{l}}{\deriv{z}} \deriv{z},
\end{equation}
where $\nu = \nu_{0} (1+z)/(1+z_{0})$, $\varepsilon(\nu,z)$ is the proper
space-averaged volume emissivity, $\tau(\nu_{0},z_{0},z)$ is the 
optical depth at frequency $\nu_{0}$ due to material along the line of sight 
from redshift $z_{0}$ to $z$ and $\deriv{l}/\deriv{z}$ is the cosmological 
line element. To solve this equation, we need to know how the emissivity 
and opacity evolve with redshift. 

\subsection{Emissivity}
For simplicity, we write the space-averaged emissivity in terms of
the global star formation rate (SFR) as
\begin{equation}
 \label{emmis}
 \varepsilon(\nu,z) = L(\nu,z) \, \dot{M}_{*} \:\: \rm{erg} \: \rm{s}^{-1} \: 
\rm{Mpc}^{-3} \: \rm{Hz}^{-1}, 
\end{equation}
where $L(\nu,z)$ is the luminosity density (i.e. the luminosity per unit 
frequency) per unit star formation rate
(in solar masses per year), and $\dot{M}_{*}$ is the global 
star formation rate, with units $\rm M_{\odot} \: yr^{-1} \: Mpc^{-3}$. 

In principle, $L(\nu,z)$ may be a complicated function of frequency 
and redshift. In practice, however, Lyman-Werner band emission is
dominated by massive, short-lived OB stars and declines rapidly once star
formation comes to an end \citep[see, for example, the instantaneous
starburst models of][in which the Lyman-Werner flux declines by an
order of magnitude within 4--$5 \: \rm{Myr}$]{lei}. Similarly, X-ray 
emission is dominated by short-lived sources such as massive X-ray binaries 
and supernova remnants which are end products of the same massive stars.
Both kinds of emission are therefore strongly correlated with the star 
formation rate; to a first approximation, we can assume that they are
directly proportional to it, and that any
redshift dependence of $L(\nu,z)$ can be neglected. With this simplification, 
determining the emissivity breaks down into two independent problems: 
determining the global star formation rate as a function of redshift, and 
determining the luminosity density as a function of the star formation rate.

\subsubsection{The star formation rate}
\label{sfr}
Although we have observational constraints on the star formation rate up to
$z \sim 5$, we have no direct constraints (and few indirect ones) at higher
redshift. Consequently, any model of high redshift star formation 
must inevitably be highly theoretical. Moreover, this lack of constraints
motivates us to choose as simple a model as possible; more complicated 
(and realistic) models can always be considered once our observational 
knowledge improves. A good example of this kind of simple model is the one 
used by \citet{har}; we adopt the same model here.

We assume that star formation proceeds primarily through starbursts, of 
duration $t_{\rm on}$ years, that are triggered when galaxies form. During 
the starburst, the star formation rate is assumed to be constant. 
The global star formation rate in this model is given by
\begin{equation}
 \dot{M}_{*} = \frac{\epsilon f_{\rm b} \Delta \rho_{\rm gal}}{t_{\rm on}} \: 
\rm{M}_{\odot} \: \rm{yr}^{-1} \: \rm{Mpc}^{-3},
\end{equation}
where $\epsilon$ is the star formation efficiency, $f_{\rm b}$ is the baryon 
fraction (ie the ratio of baryons to dark matter), and where $\Delta 
\rho_{\rm gal}$ is the cosmological density of matter in newly-formed galaxies 
(with units of $\rm{M}_{\odot} \: \rm{Mpc}^{-3}$). We assume that the value of
$f_{\rm b}$ in the protogalaxies is the same as in the IGM, or in other words
that $f_{\rm b} = \Omega_{b}/\Omega_{m}$. For the cosmological model adopted
in section~\ref{res}, this corresponds to $f_{\rm b} = 0.124$.

We further assume that the rate of change of $\Delta \rho_{\rm gal}$ is 
approximately the same as the rate of change of $F(z,T_{\rm crit})$, the 
total fraction of matter in halos with virial temperatures greater than a 
critical temperature $T_{\rm crit}$, or in other words that
\begin{equation}
 \label{diff_rho} 
\frac{\deriv{\Delta \rho_{\rm gal}}}{\deriv{z}} \simeq \rho_{m}(z) 
\frac{\deriv{F}}{\deriv{z}}(z,T_{\rm crit}), 
\end{equation}
where $\rho_{m} = 2.8 \times 10^{11} \Omega_{m} (1+z)^{3} \: \rm{M}_{\odot}
\: \rm{Mpc}^{-3}$ is the cosmological matter density. Here, $T_{\rm crit}$ 
represents the minimum virial temperature required for efficient cooling; 
to a first 
approximation, halos with $T_{\rm vir} < T_{\rm crit}$ are unable to cool, 
while those with $T_{\rm vir} > T_{\rm crit}$ cool rapidly and can form stars.
Various different definitions of $T_{\rm crit}$ are in use in the literature
\citep[see, for example][]{ro,teg,har}; we discuss our particular choice
in a later section.

Making this approximation
is equivalent to assuming that the growth in $F(z,T_{\rm crit})$ is dominated 
by the formation of new halos with $T_{\rm vir} \ge T_{\rm crit}$ (either by 
monolithic collapse or by the merger of smaller objects) rather than by the 
accretion of matter by existing halos with $T_{\rm vir} > T_{\rm crit}$. 
This is justified at high redshift when such halos are 
rare and $F(z,T_{\rm crit})$ is dominated by objects near $T_{\rm crit}$, but 
becomes less accurate at lower redshifts. For this reason (and others, to be
discussed later) we do not attempt to simulate the evolution of the background
below $z = 10$.
 
To solve equation~\ref{diff_rho} we need to know the value of $T_{\rm crit}$.
In general, this will depend both on redshift \citep[see, eg][]{teg}
and on the intensities of the UV and X-ray backgrounds. Indeed, understanding 
the evolution of $T_{\rm crit}$ with redshift is one of the main goals of 
this paper. Our procedure for
determining $T_{\rm crit}$ is discussed at length in section~\ref{meth};
for now, we assume that it is known. In this case, we can 
calculate $F(z,T_{\rm crit})$ using the Press-Schechter formalism \citep{laco}:
\begin{equation}
 F(z,M_{\rm crit}) = \rm{erfc}\left[ \frac{\delta_{c}(z)}{\sqrt{2} 
\sigma(M_{\rm crit})} \right], 
\end{equation}
where $\delta_{c}(z)$ is the critical density threshold for collapse, 
$\sigma(M)$ is the square root of the variance of the cosmological density 
field, as smoothed on a mass scale $M$, and where $M_{\rm crit}$ is the 
mass of a protogalaxy with virial temperature $T_{\rm crit}$. Although
both $\delta_{c}$ and $\sigma(M)$ depend upon the choice of cosmological 
model, their behaviour is well-known, and the problem of determining 
$F(z,M_{\rm crit})$ reduces to the relatively simple one of relating
$M_{\rm crit}$ to $T_{\rm crit}$.

To do this, we need to know the protogalactic density profile. Hydrodynamical 
simulations suggest that it is approximately isothermal \citep{abn}, but 
representing it as a singular isothermal sphere
is physically unrealistic due to the latter's infinite central density. 
Accordingly, 
we follow \citeauthor{har} and represent it as a truncated isothermal sphere 
\citep{iliev}. With this choice, we find that
\begin{equation}
\label{mt}
 M_{\rm crit} = 1.5 \times 10^{7} \left( \frac{T_{\rm crit}}{1000 \: \rm{K}}
\right)^{3/2} (1+z)^{-3/2} \Omega_{m}^{-1/2} h^{-1} \: \rm{M}_{\odot}.
\end{equation}

To determine $\Delta \rho_{\rm gal}(z)$, we must integrate 
equation~\ref{diff_rho} over a redshift interval $\Delta z(t_{\rm on})$,
corresponding to the duration of the starburst; hence,
\begin{equation}
 \label{del_rho} 
\Delta \rho_{\rm gal}(z) = \int_{z}^{z + \Delta z(t_{\rm on})}
 \rho_{m}(z) \frac{\deriv{F}}{\deriv{z}}(z,T_{\rm crit}) \, \deriv{z}.
\end{equation}

This simple model has a number of shortcomings. For instance, it assumes that
the star formation efficiency $\epsilon$ and starburst duration $t_{\rm on}$
are both constant, independent of redshift or galaxy mass. It also assumes
that each galaxy forms its stars in a single starburst and thus ignores the
effects of continuous star formation and of subsequent, merger-triggered
starbursts.~\footnote{Although both of these can be mimicked to some extent
by choosing a large value for $t_{\rm on}$.} Nevertheless, it has the virtue of
simplicity, and is a good point from which to 
start our examination of the effects of the X-ray background.

\subsubsection{The UV luminosity density}
\label{uvdensity} 
The ultraviolet flux of a star-forming galaxy is dominated by emission from
young, massive O and B-type stars. These are short-lived, with the most massive
having lifetimes of only a few Myr, and thus the ultraviolet luminosity 
density is closely correlated with the star formation rate. Its value depends 
upon the spectral properties of the newly-formed stellar population, and thus
on their initial mass function (IMF), metallicity and age.

In a recent paper, \citet{sch} presents values for the photon flux in
the Lyman-Werner bands calculated for a number of different metal-free
stellar populations. If we assume that the spectrum within the bands
is flat (a reasonable approximation), then we can
convert this photon flux into a luminosity density. For a Salpeter IMF 
with minimum mass $M_{\rm min} = 1 \: \rm{M}_{\odot}$ and maximum mass 
$M_{\rm max} = 100 \: \rm{M}_{\odot}$ \citep[model A in][]{sch}, we find
that 
\begin{equation}
\label{flux_dis}
L_{\nu} = 1.1 \times 10^{28} \:\: \rm{erg} \: \rm{s}^{-1} \: 
\rm{Hz}^{-1} \: \left(\rm{M}_{\odot} \: \rm{yr}^{-1} \right)^{-1}.
\end{equation}
Reducing $M_{\rm min}$ to the more conventional value of 
$0.1 \: {\rm M_{\odot}}$ reduces this luminosity to
\begin{equation}
\label{flux_dis2}
L_{\nu} = 4.3 \times 10^{27} \:\: \rm{erg} \: \rm{s}^{-1} \: 
\rm{Hz}^{-1} \: \left(\rm{M}_{\odot} \: \rm{yr}^{-1} \right)^{-1},
\end{equation}
as we form a greater number of low mass stars that do not contribute 
significantly to the Lyman-Werner flux.

Both of these results assume that the dissociative flux has stabilized at its
equilibrium value and is therefore proportional to the star formation rate.
This equilibrium is typically established after only 2 -- 3 Myr, so
this is generally a good approximation, even for starbursts of relatively 
short duration.

The above figures are appropriate so long as we are dealing with stars
formed out of {\em entirely} metal-free gas. Such stars are somewhat unusual, 
however, as the absence of carbon means that they are unable to generate
energy via the CNO cycle, which otherwise would dominate energy production
in stars of mass $M \simgreat 1.1 \: {\rm M}_{\odot}$. As a result, metal-free
stars are hotter than their metal-enriched counterparts \citep{ec,cc,shtu,coj}
and have harder spectra. A surprising consequence of this fact is that a
low-metallicity stellar population will produce a larger dissociative
flux than a metal-free population -- the lower effective temperatures
move the peak in the thermal emission from the most massive stars closer
to the Lyman-Werner band, causing the ionizing flux to fall but the
dissociative flux to rise.

We can use the data presented in \citet{sch} for a stellar population 
with $Z = 0.02 Z_{\odot}$ to examine the difference that this effect makes 
to the Lyman-Werner flux. Using the same IMF as in equation~\ref{flux_dis}, 
we find that 
\begin{equation}
\label{flux_ii}
L_{\nu} = 1.8 \times 10^{28}  \:\: \rm{erg} \: \rm{s}^{-1} \: 
\rm{Hz}^{-1} \: \left(\rm{M}_{\odot} \: \rm{yr}^{-1} \right)^{-1}.
\end{equation}
Thus, raising the metallicity increases the Lyman-Werner flux, but only by
about 60--70\%; as we will see in section~\ref{res_nox}, this has little
effect on the evolution of $T_{\rm crit}$.

Clearly, there are many possible models other than those considered here.
Indeed, there is growing evidence that the IMF of population III stars is
strongly biased towards high masses \citep{lar,abn,bcl,bcl2}.
However, this remains uncertain,
and in this paper we have chosen to err on the side of caution and assume
that the high-redshift IMF is similar to that at the present day.

\subsubsection{The X-ray luminosity density}
\label{def_xray}
In a recent study, \citet{helf} collate data on a number of local starburst
galaxies and compare the 2 -- 10 $\, \rm{keV}$ X-ray fluxes measured by
{\sc asca} with the 8 -- 1000 $\, \mu \rm{m}$ infrared fluxes measured by 
{\sc iras}.
They find that a clear correlation exists, with the total X-ray and infrared 
fluxes related by
\begin{equation}
\label{x_to_ir}
F_{\rm X} \simeq 10^{-4} F_{\rm IR}. 
\end{equation}
Similar correlations have
previously been reported by \citet{david} and \citet{reph} for X-rays
in the 0.5 -- 4.5 $\, \rm{keV}$ and 2 -- 30 $\, \rm{keV}$ energy bands
respectively.

Theoretically, we would expect such a correlation, with both the X-ray and
infrared emission tracing the underlying star formation rate. For the
X-rays, this occurs because the emission is dominated by massive X-ray 
binaries (MXRBs): binary systems consisting of a massive OB star accreting 
onto a compact companion (a neutron star or black hole). X-ray emission from
such systems generally switches on a few million years after the formation 
of the compact object, and the lifetime of the emitting phase is short, 
typically of the order of (2 -- 5$)\times 10^{4} \: {\rm yr}$ \citep{mv}.
These short timescales tie the emission closely to the underlying star 
formation rate \citep{wg}. The far-infrared flux, on the other hand, 
tracks star formation far more directly, being dominated by emission from dust 
heated by short-lived, massive stars. 

To use this observed correlation to determine the X-ray luminosity of a 
star-forming galaxy as a function of its star formation rate, we use the 
result from the starburst models of \citet{lh} that 
\begin{equation}
\label{ir}
L_{\rm IR} \sim 1.5 \times 10^{10} \left( \frac{\rm{SFR}}{1 \: \rm{M}_{\odot}
\: \rm{yr}^{-1}} \right) \: \rm{L}_{\odot}, 
\end{equation} 
together with equation~\ref{x_to_ir} to write the X-ray luminosity as
\citep{helf}
\begin{equation}
\label{lx}
 L_{\rm X} = 6 \times 10^{39} \left( \frac{\rm{SFR}}{1 \: \rm{M}_{\odot}
\: \rm{yr}^{-1}} \right) \: \rm{erg} \: \rm{s}^{-1}.
\end{equation} 
We then assume that this relationship between X-ray luminosity and star
formation rate remains valid as we move to higher redshifts. Note that
the same need not be true for the relationship between X-ray luminosity
and infrared luminosity, or infrared luminosity and star formation rate:
although they serve to establish the correlation between X-ray luminosity
and star formation rate at $z = 0$, the assumption that this correlation
remains valid at higher redshift does not imply that these other correlations
also remain valid. Indeed, we would expect the infrared luminosity of 
dust-free protogalaxies to be very much lower than would be predicted by 
equation~\ref{ir}. 

Evidence that the correlation between X-ray luminosity and star formation 
rate does indeed remain valid at high redshift is provided by the recent 
stacking 
analysis of individually undetected Lyman break galaxies in the {\em Chandra} 
Deep Field-North \citep{cdfn}. This analysis finds that the average rest frame 
luminosity of the Lyman break galaxies in the 2 -- 8 $\: \rm{keV}$ energy
band is $L_{\rm X} = 3.2 \times 10^{41} \: \rm{erg} \: \rm{s}^{-1}$. 
Assuming a typical star formation rate of $50 \: \rm{M_{\odot}} \: 
\rm{yr}^{-1}$  for these galaxies \citep{shst}, this corresponds to an
X-ray luminosity of  
\begin{equation}
 \label{lbg_xray} 
 L_{\rm X} = 6.4 \times 10^{39} \left( \frac{\rm{SFR}}{1 \: \rm{M}_{\odot}
\: \rm{yr}^{-1}} \right) \: \rm{erg} \: \rm{s}^{-1},
\end{equation} 
consistent with the value derived above.~\footnote{Note that X-ray luminosity
of equation~\ref{lbg_xray} is measured in a slightly different energy band
from that of equation~\ref{lx}, so the agreement between the two values is
not quite as good as may at first appear. Nevertheless, the necessary 
correction is small, and the values agree to within 50\%.}
Although far from conclusive, this result suggests that we can extrapolate
the locally observed correlation to at least as far as $z \simeq 4$.
 
Comparing our determination of $L_{\rm X}$ with a recent calculation by
\citet{oh}, we find a difference of a factor of ten in our results.
Some of this disagreement is due to the difference in X-ray energy bands 
considered ($0.2$ -- $10 \: \rm{keV}$ in \citet{oh}, compared to $2$ -- $10 
\: \rm{keV}$ here), but some must surely be due to intrinsic scatter in the 
observational data, suggesting that equation~\ref{lx} should  properly be 
regarded as an order of magnitude estimate of the true X-ray luminosity.

Given equation~\ref{lx} for the X-ray luminosity as a function of the star
formation rate, we calculate the X-ray luminosity density by assuming 
a template spectrum of power-law form
\begin{equation}
 L_{\nu} = L_{0} \left( \frac{\nu_{0}}{\nu} \right)^{\alpha},
\end{equation}
where $h \nu_{0} = 1 \: {\rm keV}$, and requiring that
\begin{equation}
 L_{\rm X} = \int_{\nu_{1}}^{\nu_{2}}  L_{\nu} \, \deriv{\nu}, 
\end{equation} 
where $h \nu_{1} = 2 \: {\rm keV}$ and $h \nu_{2} = 10 \: {\rm keV}$.

\citet{reph} find that a weighted average of the galaxies in their
sample gives a value for the spectral index of $\alpha = 1.5  \pm 0.3$.
Adopting this value and solving for $L_{0}$, we find that
\begin{equation}
 L_{0} = 3.4 \times 10^{22} \:\: \rm{erg} \: \rm{s}^{-1} \: 
\rm{Hz}^{-1} \: \left(\rm{M}_{\odot} \: \rm{yr}^{-1} \right)^{-1}.
\end{equation}
Altering $\alpha$ changes $L_{0}$, but never by more than 50\%
for values consistent with the  \citeauthor{reph} measurement.
Clearly, individual galaxies may have spectra that differ markedly from this 
simple template, but it should be a reasonable approximation when averaging
over a large number of galaxies.

\subsubsection{Alternative X-ray models} 
\label{alt_xray} 
The above model is simple, and empirically motivated, but does 
assume that the X-ray emission of high-redshift star-forming galaxies 
is very similar to that of starbursts observed locally. This is a reasonable
assumption in the absence of evidence to the contrary, and is probably valid
as long as massive X-ray binaries continue to dominate the galactic X-ray
emission. However, it is quite possible that at high redshift some other type
of source will come to dominate the emission, particularly if the number of 
binary systems is small, as is suggested by recent simulations \citep{abn}. 
It is therefore prudent to consider the effects of other potential sources 
of X-rays. 

The obvious candidates are supernova remnants (SNR); next to X-ray 
binaries, they are the most significant galactic sources \citep{helf}.
They can emit X-rays through a variety of different emission mechanisms, 
but at high redshift the most significant will be thermal bremsstrahlung 
emission and non-thermal inverse Compton emission.

Thermal bremsstrahlung is produced by the hot gas within the SNR.
Detailed modeling properly requires a hydrodynamical treatment 
\citep[see, for example,][]{chev}, but for our purposes a simple 
parameterization suffices. If we assume that the hot gas has a single
characteristic temperature $T_{\rm x}$, then we can write the luminosity
density per unit star formation rate as
\begin{equation}
 L_{\nu} = L_{\rm b} \left(\frac{h \nu_{0}}{kT_{\rm x}} \right) 
 \exp \left(\frac{-h\nu}{kT_{\rm x}} \right)
 \rm{erg} \: \rm{s}^{-1} \: \rm{Hz}^{-1} \: \left(\rm{M}_{\odot} \: 
 \rm{yr}^{-1} \right)^{-1}.
\end{equation}
where $h\nu_{0} = 1 \: \rm{keV}$ and where $L_{\rm b}$ is constant.
Moreover, we can write the total X-ray luminosity as
\begin{equation}
 L_{\rm tot} = 3.2 \times 10^{43} f_{\rm x} E_{51} N_{\rm sn} \:\,
  \rm{erg} \: \rm{s}^{-1} \, \left(\rm{M}_{\odot} \: 
  \rm{yr}^{-1} \right)^{-1},
\end{equation} 
where $E_{51}$ is the typical supernova explosion energy (in units of 
$10^{51} \: \rm{ergs}$), $f_{\rm x}$ is the fraction of this energy radiated 
as bremsstrahlung, and where $N_{\rm sn}$ is the number of supernovae that
explode per solar mass of stars formed. The value of $N_{\rm sn}$ depends 
on the IMF; for the standard Salpeter IMF adopted previously, $N_{\rm sn} 
= 0.0075 \: {\rm M}^{-1}_{\odot}$. Finally, since
\begin{equation}
 L_{\rm tot} = \int_{0}^{\infty}  L_{\nu} \, \deriv{\nu}, 
\end{equation}
we can fix the value of $L_{\rm b}$; we find that
\begin{equation}
 L_{\rm b} = 1.3 \times 10^{26} f_{\rm x} E_{51} N_{\rm sn}  
 \: \rm{erg} \: \rm{s}^{-1} \: 
 \rm{Hz}^{-1} \left(\rm{M}_{\odot} \: \rm{yr}^{-1} \right)^{-1}.
\end{equation}

For typical supernova parameters ($E_{51} = 1$ and an ambient density 
$n = 1 \: \rm{cm}^{-3}$), \citet{helf} find that a fraction 
$f_{\rm x} = 2 \times 10^{-4}$  of the explosion energy is radiated, at a 
characteristic temperature $kT_{\rm x} = 1 \: \rm{keV}$. 
On the other hand, the higher mean density at high redshift, together with
the comparative weakness of outflows from low metallicity stars \citep{kud}
suggest that the typical ambient density may be very much higher.
In particular, if it is as high as $n \simeq 10^{7} \: \rm{cm}^{-3}$, then
a supernova remnant will radiate its energy extremely rapidly, before the
ejecta have time to thermalize \citep{csnr}. In this case, the fraction of
energy radiated as bremsstrahlung is very much higher 
($f_{\rm x} \simeq 0.01$),
as is the characteristic temperature ($kT_{\rm x} \sim 30 \: \rm{keV}$).
We examine both of these models in section~\ref{res}, with the understanding 
that the true picture lies somewhere in between.

In addition to this thermal emission, supernova remnants also produce 
non-thermal X-rays. These are generated as the relativistic electrons
produced by the SNR gradually lose energy through synchrotron radiation,
non-thermal bremsstrahlung and the inverse Compton scattering of 
photons from the cosmic microwave background. At high redshift, the latter
is likely to dominate \citep{oh}. The spectrum of the resulting
emission depends upon the energy spectrum of the relativistic electrons, 
but at the energies of interest is well represented by a power law:
$L_{\nu} \propto \nu^{-1}$. The intensity of the emission depends upon the 
fraction of the supernova energy transferred to the electrons; this is not
well constrained, with estimates ranging from 0.1\% to 10\%. Accordingly, 
we model the emission as 
\begin{equation}
 L_{\nu} = 7.7 \times 10^{23} \left( \frac{\nu_{0}}{\nu} \right) f_{\rm e} 
\: \rm{erg} \: \rm{s}^{-1} \: 
\rm{Hz}^{-1} \left( M_{\odot} \: \rm{yr}^{-1} \right)^{-1},
\end{equation}
where $h\nu_{0} = 1 \: \rm{keV}$, $f_{\rm e}$ is the fraction of energy 
deposited in the electrons, and where we have assumed that $E_{51} = 1$ 
and $N_{\rm sn} = 0.0075 \: {\rm M}^{-1}_{\odot}$ as in the 
thermal bremsstrahlung case. This expression assumes a high-energy cutoff 
for the X-ray spectrum at $10 \: \rm{keV}$ but is only logarithmically 
dependent on the value of this cutoff. In section~\ref{res}, we examine 
results for models with $f_{\rm e} = 10^{-3}$ and $f_{\rm e} = 0.1$,
which bracket the range of plausible values.

Much more detail on high redshift inverse Compton emission, including a 
discussion of potential observational tests, is given in \citet{oh}.

\subsection{Opacity}
The opacity $\tau(\nu_{0},z_{0},z)$ can be separated into two distinct 
components -- absorption by dust and gas within the emitting protogalaxy,
which we term intrinsic absorption, and absorption by gas along the line
of sight through the IGM.

Intrinsic absorption is difficult to model with any degree of accuracy as
it depends upon a number of variables -- the size and shape of the galaxy,
 its ionization state, the position of the sources within it, 
the dust content etc. Rather than attempt to model these in detail
-- a significant undertaking in itself -- we instead adopt a highly 
approximate representation. We assume that the emitted X-rays are attenuated 
by absorption by a neutral hydrogen column density of $N_{\mH} = 10^{21} 
\: \rm{cm}^{-2}$ plus an associated neutral helium column density
$N_{\He} = 0.08 N_{\mH}$. This absorption is assumed to be the same 
for all sources. These values are chosen because they are representative of 
the column densities of the protogalaxies studied in this paper (which 
presumably contain the bulk of the X-ray sources). Reducing $N_{\mH}$
(as would be appropriate if much of the surrounding \hi were photoionized
and/or dispersed by the progenitors of the X-ray sources) has little or no 
effect on $T_{\rm crit}$, as gas in the IGM and within the protogalaxy itself
quickly come to dominate the total absorption. Increasing $N_{\mH}$, on the 
other hand, has more significant effects: an order of magnitude increase in 
$N_{\mH}$ produces similar results to an order of magnitude decrease in the 
strength of the X-ray background, which, as we will see in section~\ref{res},
is generally sufficient to render X-ray feedback ineffective. Consequently,
we will overestimate the effect of the background if the bulk of the X-ray
sources reside within massive galaxies.

Turning to the Lyman-Werner bands, we note that intrinsic absorption will
generally be negligible within small protogalaxies, as their $\mHt$ content 
is rapidly photodissociated \citep{scog}. Moreover, we also assume that
the effects of dust absorption are negligible. In galaxies of primordial 
composition, this is obviously true; in metal-poor galaxies, it should also
be a good approximation, as very large column densities are required for
significant dust obscuration (for instance, $N_{\mH} \simgreat 10^{23} \: 
\rm{cm}^{-2}$ for $Z = 10^{-2} Z_{\odot}$ gas, if the dust-to-gas ratio is
similar to that in the Milky Way). Again, these assumptions break down if
the majority of sources are to be found in massive, metal-rich galaxies,
but we expect such galaxies to be extremely rare at the redshifts of interest
in this paper.

Compared to intrinsic absorption, the effects of absorption due to the IGM
are much simpler to treat, particularly if we can assume that the bulk of the 
gas remains at an approximately uniform density. This assumption proves 
reasonable at high redshift for photons with mean free paths much greater 
than the typical clumping scale, as is the case for both Lyman-Werner band 
photons and X-rays. Our treatment of IGM absorption is discussed in the 
following sections.

\subsubsection{Lyman-Werner band absorption}
\label{lwabs}
The continuum opacity of metal-free gas is very small \citep{lcs} and for 
our purposes can be neglected. Consequently, the only significant sources
of opacity encountered by Lyman-Werner photons are absorption by the Lyman 
series lines of neutral hydrogen, and by the Lyman-Werner lines of molecular
 hydrogen. 
%%\begin{figure}
%%\centering
%%\epsfig{figure=plots/sawtooth.eps,width=5.8cm,angle=270,clip=}
%%\caption[The `sawtooth' spectrum of the Lyman-Werner background]{An example
%%of the `sawtooth' pattern that neutral hydrogen absorption creates in the 
%%spectrum of the Lyman-Werner background. This example shows the background 
%%at $z=30$, and is taken from a model with star formation efficiency 
%%$\epsilon=0.1$ and starburst lifetime $t_{\rm on}=10^{7} \: \rm{yr}$.}
%%\label{saw}
%%\end{figure}
The Lyman series lines have the effect of absorbing any Lyman-Werner photons
of the same frequency, and reprocessing them to Lyman-$\alpha$ photons plus
associated softer photons. As  Lyman-$\alpha$ lies outside of the Lyman-Werner
 band, the net effect is to block from view any sources at redshifts higher
than some maximum, $z_{\rm max}$, given by
\begin{equation}
 \frac{1 + z_{\rm max}}{1+ z_{0}} = \frac{\nu_{\rm H}}{\nu_{0}},
\end{equation}
where $\nu_{\rm H}$ is the frequency of the appropriate Lyman series line
and $\nu_{0}$ and $z_{0}$ are the observed frequency and redshift.
Clearly, the size of $z_{\rm max}$ depends upon the distance between $\nu_{0}$
and $\nu_{\rm H}$, and thus more sources are seen at frequencies that are
a long way from a line. As a result, the spectrum develops a characteristic
`sawtooth' shape (see figure 1 in \citealt{har}), with the effect becoming 
more pronounced as one nears the Lyman limit.

Absorption by molecular hydrogen is rather more complicated, 
due to the large number of Lyman-Werner lines that  contribute 
to the opacity. If we approximate the lines as delta functions then an
individual line produces an opacity
\begin{equation}
 \tau_{i} = \frac{\pi e^{2}}{m_{\rm e}c} f_{{\rm osc},i} \lambda_{i} 
\frac{n_{\mHt,{\it i}}(z_{i})}{H(z_{i})},
\end{equation}
where $f_{{\rm osc},i}$ and $\lambda_{i}$ are the oscillator strength and
wavelength of the transition, $n_{\mHt,{\it i}}(z_{i})$ is the number density 
of $\mHt$ molecules in the level giving rise to the line at $z_{i}$, the
redshift of absorption, and $H(z_{i})$ is the Hubble constant at $z_{i}$. 
The value of $z_{i}$ is given by
\begin{equation}
\frac{1+z_{i}}{1+z_{0}} = \frac{\nu_{i}}{\nu_{0}},  
\end{equation} 
where $\nu_{i} = c / \lambda_{i}$.

If we assume that all of the photons that are absorbed in the lines are
permanently removed from the Lyman-Werner band, then the total opacity
$\tau(\nu_{0},z_{0},z)$ is simply given by the sum over all lines with
$z < z_{i} < z_{\rm max}$:
\begin{equation}
\tau(\nu_{0},z_{0},z) = \sum_{i} \frac{\pi e^{2}}{m_{\rm e}c} 
f_{{\rm osc},i} \lambda_{i} \frac{n_{\mHt,{\it i}}(z_{i})}{H(z_{i})}.
\end{equation}
This sum potentially involves a very large number of lines, but can be greatly 
simplified by assuming that all of the $\mHt$ is to be found in its ortho or
para ground state; at the redshifts of interest, the population of excited
states will be negligible. 

In deriving this expression, we have assumed that every absorption permanently
removes a Lyman-Werner band photon. This is not entirely 
correct. On average, only 15\% of absorptions are followed by photodissociation
of the $\mHt$ molecule \citep{db}; the rest of the time, the molecule decays
back to a bound state, emitting a photon. In their treatment of this
problem, \citet{har} assumed that the excited $\mHt$ molecule would always
decay directly back into the original state, and would thus emit a photon of
the same energy as the one initially absorbed. In fact, this is not correct 
(T.\ Abel, private communication); most decays occur initially to highly 
excited vibrational states, producing photons redwards of the Lyman-Werner 
bands. Only a small fraction of decays (about 5\%) take place directly into 
the original state, while a slightly larger fraction (about 15\%) produce 
photons that lie elsewhere in the Lyman-Werner band system \citep{scog}. 
We do not include the effect of these photons, however; an accurate treatment
would be quite complicated and is almost certainly unnecessary -- as we shall 
see in section~\ref{res}, $\mHt$ in the IGM is rapidly destroyed by the 
growing Lyman-Werner background and is completely negligible by the time
that negative feedback begins.

\subsubsection{X-ray absorption}
At X-ray energies, the opacity of the intergalactic gas is dominated by 
the ionization of neutral 
hydrogen and helium; prior to recombination, the $\Hep$ abundance is 
small and can be neglected. The X-ray opacity can thus be written as
\begin{equation}
 \label{tauigm}
 \tau(\nu_{0},z_{0},z) = \int^{z}_{z_{0}} 
\left[ \sigma_{\mH}(\nu) \, n_{\mH} + \sigma_{\He}(\nu) \, n_{\He} \right]
\frac{\deriv{l}}{\deriv{z}} \deriv{z}, 
\end{equation}
where
\begin{equation}
 \nu = \nu_{0} \frac{(1+z)}{(1+z_{0})},
\end{equation}
and where $\sigma_{\mH}(\nu)$ and $\sigma_{\He}(\nu)$ are the absorption 
cross-sections of neutral hydrogen and helium respectively, with
$n_{\mH}(z)$ and $n_{\He}(z)$ being the corresponding number densities.
As long as the fractional ionization of the IGM remains small (i.e. a few
percent or less), the ratio between $n_{\He}$ to $n_{\mH}$ can be
accurately approximated by its primordial value
\begin{equation}
 \frac{n_{\He}}{n_{\mH}} = \frac{Y}{4 - 4Y}
\end{equation}
where $Y=0.247$ is the helium mass fraction, and we can write 
equation~\ref{tauigm} purely in terms of $n_{\mH}$ as
\begin{equation}
 \label{tauigm2}
 \tau(\nu_{0},z_{0},z) = \int^{z}_{z_{0}} \left[\sigma_{\mH} + \left( 
\frac{Y}{4-4Y} \right) \sigma_{\He} \right] n_{\mH} 
\frac{\deriv{l}}{\deriv{z}} \deriv{z}.
\end{equation}
This integral is readily computable by means of numerical integration.

\section{Method}
\label{meth}
In the previous section, we showed that, given a simple star formation model,
it is relatively easy to calculate the evolution of the Lyman-Werner and 
X-ray backgrounds. Two of the parameters of our star formation model -- the
star formation efficiency $\epsilon$ and starburst lifetime $t_{\rm on}$
-- we treat as free parameters (although they can be constrained to some
 extent -- see, e.g. \citealt{hl}). The remaining parameter, $T_{\rm crit}$,
the temperature at which efficient cooling becomes possible, is determined
by the strength of the backgrounds themselves. 
This clearly presents us with a problem: the evolution of $T_{\rm crit}$
is coupled to that of the backgrounds, and to know one we must first
know the other. 

Fortunately, this difficulty is easily avoided. We know that at high redshift 
the number of protogalaxies, and hence the star formation rate, must be very 
small. Consequently, there must be some redshift above which the external
radiation field will become too weak to affect galaxy formation. The precise 
redshift at which this occurs is model dependent, but for the models examined 
in this paper we typically find that radiative feedback is negligible above
$z = 40$. By choosing an initial redshift $z_{i} = 50$, therefore, we can be 
sure that in our initial simulation the background radiation will have no 
effect.

Given this starting point, we next proceed incrementally to lower redshifts
via the following procedure:
\begin{enumerate}
\item Given $T_{\rm crit}(z_{i})$, we calculate the background radiation field
at $z = z_{i} - \Delta z$, assuming that $T_{\rm crit}(z) = 
T_{\rm crit}(z_{i})$.

\item Using the computed background, we simulate the chemical and thermal 
evolution of a protogalaxy with $T_{\rm vir} = T_{\rm crit}(z)$; the details
of this simulation are outlined in sections~\ref{comp} to \ref{halt} below. 
The main aim of this
simulation is to determine whether the protogalactic gas can cool efficiently.

\item If the protogalactic gas cools, then our assumed value of $T_{\rm crit}$
is correct; we store this result, and return to step one to proceed to the next
redshift. If the gas fails to cool, we continue to step four.

\item We increment our assumed value of $T_{\rm crit}(z)$ by a small
amount $\Delta T$, and recalculate the background radiation field. 
We assume that $T_{\rm crit}$ varies linearly over $\Delta z$. Given the new 
background, we return to step two.

\end{enumerate}
Provided that $\Delta z$ and $\Delta T$ are both small, the error in
 $T_{\rm crit}(z)$ will also be small; this is particularly the case once
emission from larger protogalaxies (which cool via Lyman-$\alpha$ radiation) 
begins to dominate the background.

This approach reduces the coupled problem to the simpler one of determining
whether a protogalaxy with virial temperature $T_{\rm vir}$ and formation 
redshift $z_{f}$ will cool when exposed to a particular background radiation 
field.
To answer this question, we need to be able to model the thermal and chemical
evolution of the protogalaxy. Our approach to this problem is outlined in the 
following sections.

\subsection{Computing the evolution of protogalactic gas}
\label{comp}
Ideally, we would like to use a high-resolution hydrodynamical simulation
to follow the thermal and chemical evolution of the protogalactic gas
\citep[see, e.g.][]{mba}. Unfortunately, including the effects of radiative
transfer, particularly of photons in the Lyman-Werner bands, into such a 
simulation is not currently feasible. We are thus forced to approximate.
In choosing an appropriate approximation, we are also motivated by the 
desire to minimize the computational requirements of our simulations, 
so that we can explore the effects of a variety of different source models.
We make three main approximations:
\begin{enumerate}
\item We assume spherical symmetry. This is a reasonable approximation for
the first generation of protogalaxies, but clearly is incorrect in detail
\citep[see, e.g.\ figure 2 of ][]{abn}.

\item We assume that the protogalactic gas is {\em static}, at least on the
timescale of the simulation. This allows us to ignore the hydrodynamical 
evolution of the gas, and also substantially simplifies the treatment of  
radiative transfer. This assumption clearly breaks down once the gas begins to
cool strongly and loses its pressure support, but as we are only interested in
the evolution up to this point, this is not a significant problem.

\item We assume that all of the $\mHt$ molecules remain in the rotational
and vibrational ground state, in either ortho or para form. This 
simplification allows us to evolve the chemistry and radiative transfer
on the timescale on which the total $\mHt$ abundance changes 
(typically $10^{11}$ -- $10^{12} \: \rm{s}$)
rather than that on which the individual level populations change 
($10^{6}$ -- $10^{10} \: \rm{s}$). It also simplifies our treatment of the
radiative transfer. We discuss this approximation in more detail in 
section~\ref{pd}.
\end{enumerate}

Together, these approximations allow us to solve for the chemical and thermal
evolution of a model protogalaxy in a matter of minutes on a fast desktop
computer. This allows us to study the redshift evolution of $T_{\rm crit}$ 
at high resolution in both temperature and redshift and for a number of 
different X-ray source models.

However, this approach has an obvious drawback -- we cannot be sure that 
our approximations give a fair representation of the real protogalaxy. Of
particular concern is the neglect of the hydrodynamic evolution of the gas,
and the consequent error in the density profile. This is potentially 
significant because the $\mHt$ cooling rate, along with many of the chemical 
reaction rates, scales as the square of the density. Small errors in the 
density can thus lead to larger ones in the computed temperature. These 
concerns are mitigated to some extent, however, by the close agreement 
between the
results of detailed numerical simulations and previous semi-analytic 
treatments. For instance, the values of $T_{\rm crit}$ obtained from 
the smoothed-particle hydrodynamics simulations of \citet{fc} agree well
with the results of \citet{teg}, despite the highly approximate 
uniform density profile adopted by the latter group. Similarly, the results
of \citet{mba}, obtained with a three-dimensional adaptive mesh hydrodynamical
code broadly agree with those of \citet{har}, who use a static model similar 
to that presented here. Together, these results suggest that $T_{\rm crit}$ is
insensitive to the precise details of the density profile, but clearly
this remains an area of concern.
 
Our computational method can be broken down into three main stages --
initialization of the density profile and the chemical abundances, 
computation of the thermal and chemical evolution of the gas, and 
termination of the simulation at a suitable point. These are described below.

\subsection{The model protogalaxy}
\label{model_pg}
The protogalactic density profile is modeled as a truncated isothermal sphere,
with central overdensity $\delta = 1.796 \times 10^{4}$ and truncation radius
\begin{equation}
 r_{t} = 4.62 \times 10^{3} \left( \frac{T_{\rm vir}}{1000 \: \rm{K}}
\right)^{1/2} (1+z_{\rm f})^{-3/2} \Omega_{m}^{-1/2} h^{-1} \: \rm{pc}.
\end{equation}
The virial temperature and redshift of formation of the protogalaxy completely
specify its density profile.

We subdivide this profile into $N_{g}$ spherical shells of uniform 
thickness and compute the mean density within each shell. We have run a 
number of test simulations with different values of $N_{g}$, and find that
setting $N_{g} = 100$ provides sufficient spatial resolution to accurately 
determine $T_{\rm crit}$.

We assume that the initial chemical composition of the protogalactic gas is the
same as that of the intergalactic medium. At our initial redshift $(z=50)$, 
we take this from \citet{sld}. At lower redshifts, the chemical
evolution of the IGM is influenced by the Lyman-Werner and X-ray backgrounds. 
We therefore calculate the intergalactic abundances explicitly, using the
chemical model outlined in section~\ref{chem}, by solving the chemical 
rate equations:
\begin{equation}
 \frac{\deriv{n_{i}}}{\deriv{t}} = C_{i}(n_{j},T) - D_{i}(n_{j},T)n_{i}, 
\end{equation}
where $C_{i}$ and $D_{i}$ are source and sink terms for $n_{i}$.
At the same time we also solve for the temperature of the intergalactic gas
\citep{gp}
\begin{equation}
 \frac{\deriv{T}}{\deriv{t}} = - 2TH(z) + \frac{2}{3 k n_{\rm tot}} 
\left( \Gamma - \Lambda \right),
\end{equation}
where $H(z)$ is the Hubble constant, $n_{\rm tot}$ is the total particle
number density and where $\Gamma$ and $\Lambda$ are the net heating and 
cooling rates (see section~\ref{cool}). 

We solve this set of ordinary differential equations with the {\sc stifbs}
integrator of \citet{nr}. As the size of the required timestep is 
generally much smaller than the redshift interval $\Delta z$ that separates
our individual determinations of $T_{\rm crit}$, we compute intermediate
values by interpolation and from these determine the strength of the radiation
background and hence the photochemical rates. Although our main aim in 
following this chemistry is to determine the correct initial abundances for
our simulations of protogalactic evolution, the results are of interest in
their own right and are presented and discussed in section~\ref{ther_chem}.

Our treatment of the chemistry of the IGM does not include the effects of 
the ionizing photons from stars (and/or quasars) that are ultimately 
responsible for the
reionization of the intergalactic gas. This is justified at early epochs, as
these photons are confined within small \hii regions
surrounding the luminous sources, but this simplification restricts the 
validity of our results to the period prior to cosmological reionization. 
The post-reionization epoch, and the effect of reionization on galaxy 
formation, have received extensive study elsewhere \citep[see][and references 
therein]{itb}.

\subsection{The chemical model}
\label{chem}
To simulate the chemical evolution of the protogalactic gas, we adopt a
chemical model consisting of thirty reactions between nine species:
$\mH$, $\Hm$, $\Hp$, $\He$, $\Hep$, $\Hepp$, $\mHt$, $\mHtp$ and free 
electrons. The reactions included in the model are summarized in 
table~\ref{chemtab}, 
together with the source(s) of the data used. This model is based in
large part on that of \citet{aanz}, but has been modified to improve its
accuracy when applied to optically thick gas. Aside from a number of updates
to the reaction coefficients in the light of new data, the main differences
are as follows:

\begin{enumerate}

\item We include the contribution to the hydrogen ionization rate arising from
the ionizing photons produced by $\Hep$ recombination, in line with the
discussion in chapter 2 of \citet{os}. Although commonly a small correction
to the total rate, this can become significant when X-ray photoionization
dominates.

\item To enable us to accurately determine the $\Hep$ abundance, we find 
that we need to include the effects of charge transfer between $\Hep$ and 
$\mH$ (reaction 20), as this can be comparable to the recombination rate 
when the fractional ionization is small. For completeness we also include
the inverse reaction (no.\ 21), although this is unimportant at 
$T < 10^{4} \: \rm{K}$.

\item We include the contribution to the ionization rates of hydrogen and
helium arising from secondary ionization by energetic photoelectrons, 
based on the recent calculations of \citet{dyl}. The contribution of 
secondary ionization to the other photoionization rates is small and can be
neglected.

\item We do not include the photodissociation of $\mHt$ by photons above the
Lyman limit (reaction 28 in \citeauthor{aanz}), as in optically thick gas
this will be negligible compared to the effects of $\mHt$ photoionization
(reaction 26). On the other hand, we do include the effects of dissociative 
photoionization (reaction 27), which becomes significant for photon energies
greater than 30$\: \rm{eV}$.

\end{enumerate}

More information about all of these points, and the chemical model generally,
can be found in \citet{scogPHD}.

\begin{table}
\begin{tabular}{lll}
\hline
\multicolumn{1}{c}{No.} & \multicolumn{1}{l}{Reaction} & 
\multicolumn{1}{c}{Reference} \\
\hline
1. & $\mH + \me \rightarrow \Hp + 2 \me$ & \citet{janev} \\
2. & $\Hp + \me \rightarrow \mH + \gamma$ & \citet{fph} \\
3. & $\He + \me \rightarrow \Hep + 2 \me$ & \citet{janev} \\
4. & $\Hep + \me \rightarrow \Hepp + 2 \me$ & \citet{aanz} \\
5. & $\Hep + \me \rightarrow \He + \gamma $ & \citet{hs}, \\ 
 & & Aldrovandi \& \\
 & & \hspace{10pt} Pequignot~(\citeyear{ap}) \\
6. & $\Hepp + \me \rightarrow \Hep + \gamma $ & \citet{fph} \\
7. & $\mH + \me  \rightarrow  \Hm + \gamma$ & \citet{wish} \\
8. & $\Hm + \mH \rightarrow \mHt + \me$ & \citet{ldz} \\
9. & $\mH + \Hp \rightarrow \mHtp + \gamma$ & \citet{rp} \\
10. & $\mHtp + \mH \rightarrow \mHt + \Hp$ & \citet{kah} \\
11. & $\mHt + \Hp \rightarrow  \mHtp + \mH$ & \citet{hmf} \\
12. & $\mHt + \me \rightarrow 2\mH + \me$ & \citet{st} \\
13. & $\mHt + \mH \rightarrow 3\mH$ & \citet{msm} \\
14. & $\Hm + \me  \rightarrow  \mH + 2\me$ & \citet{janev} \\
15. & $\Hm + \mH \rightarrow  2\mH + \me$ & \citet{janev} \\
16. & $\Hm + \Hp \rightarrow 2\mH $ & \citet{map} \\
17. & $\Hm + \Hp \rightarrow \mHtp + \me$ & \citet{po} \\
18. & $\mHtp + \me \rightarrow 2\mH$ & \citet{sdgr} \\
19. & $\Hm + \He \rightarrow \mH + \He + \me$ & \citet{huqhe} \\
20. & $\Hep + \mH \rightarrow \He + \Hp + \gamma $ & \citet{zyg} \\
21. & $\He + \Hp \rightarrow \Hep + \mH$ & \citet{kim} \\
22. & $\mH + \gamma \rightarrow \Hp + \me$ & \citet{os} \\
23. & $\He + \gamma \rightarrow \Hep + \me$ & \citet{ysd} \\
24. & $\Hep + \gamma \rightarrow \Hepp + \me$ & \citet{os} \\
25. & $\Hm + \gamma \rightarrow \mH + \me$ & \citet{dj} \\
26. & $\mHt + \gamma \rightarrow \mHtp + \me$ & \citet{or}, \\ 
 & & \citet{wam}, \\
 & & \citet{ysd} \\
27. & $\mHt + \gamma \rightarrow \mH + \Hp + \me$ & \citet{sh} \\
28. & $\mHtp + \gamma \rightarrow \mH + \Hp$ & \citet{dunn} \\
29. & $\mHtp + \gamma \rightarrow 2\Hp + \me$ & \citet{bo} \\
30. & $\mHt + \gamma \rightarrow \mHt^{*} \rightarrow 2\mH$ & See text \\
\hline
\end{tabular}
\caption{A list of the reactions included in our chemical model of 
protogalactic gas. Values for the rate coefficients (or radiative
cross-sections where appropriate) are given in \citet{scogPHD}.
References are to the primary source(s) of the data whenever possible;
in many cases, we have also used analytical fits to this data from
\citet{aanz}, \citet{gp} or \citet{sld}.}
\label{chemtab}
\end{table}

\subsection{Radiative transfer}
\label{rtmethod}
In order to determine rate coefficients for the photochemical reactions
(nos.\ 22--30), we need to calculate the evolution of the mean specific 
intensity $J_{\nu}$ within the protogalactic gas. For optically thin gas,
this is simple, but the only photochemical reactions for 
which we can assume optically thin conditions are $\Hm$ photodetachment 
(reaction 25) and $\mHtp$ photodissociation (reaction 28); for the rest,
we must supplement our chemical model with a treatment of radiative transfer. 

The high degree of symmetry of the problem makes direct solution of the
radiative transfer equation feasible, but the large spectral resolution
required to properly treat the Lyman-Werner lines 
($\Delta \nu/\nu \simeq 10^{-5}$) makes such an approach very time-consuming,
particularly since it must be repeated for a very large number of timesteps.
This motivates our choice of a much simpler approximate treatment that 
nevertheless should produce results of an acceptable accuracy. 

To begin with, we assume that the re-emission and/or scattering of photons 
within the protogalactic gas is negligible. Although highly accurate for the
X-ray photons, this approximation is less so for the Lyman-Werner photons,
since many Lyman-Werner absorptions are followed by re-emission. However,
as noted previously in section~\ref{lwabs}, the majority of these photons are 
emitted at frequencies that do not coincide with the ground-state Lyman-Werner
lines, and rapidly escape from the protogalaxy without being re-absorbed.
By ignoring the small fraction of re-emitted photons that do coincide with
ground-state lines, we underestimate the $\mHt$ photodissociation rate 
slightly, but this effect is small compared to the uncertainty in the incident
spectrum.

In addition to the ionizations produced by the incident X-rays (and secondary
photoelectrons), we also must account for the ionizations caused by the
emission of ionizing photons during $\mH$ and $\He$ recombination within the
gas. This is done by use of the on-the-spot approximation, as outlined
in \citet{os}. This approximation is valid as long as the mean free path of 
the ionizing photons is much shorter than any length scale of interest. For
the protogalaxies that we study, this condition is generally satisfied;
for example, the mean free path of a photon just above the Lyman limit in the 
core region of a $z=10$ protogalaxy is only $0.01 \: \rm{pc}$ (and is even 
smaller in higher redshift objects). It breaks down somewhat in the outermost
regions of our protogalaxies, where the density is several orders of 
magnitude lower, but $\mHt$ formation in these regions is negligible. 

\begin{figure}
\centering
\epsfig{figure=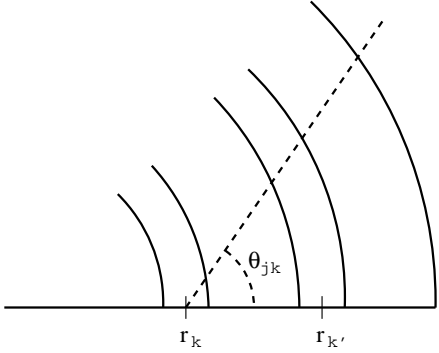,width=5.8cm,angle=0,clip=}
\caption{A schematic diagram of the radiative transfer geometry used in our
simulations. For each of the $N_{g}$ mass shells, we pick a single central 
point ($r_{k}$ and $r_{k'}$ above are two examples) and compute column 
densities along $N_{a}$ lines of sight to this point. With each line of 
sight, we can associate an angle; for instance, with the $j$-th line of sight 
from $r_{k}$, we associate the angle $\theta_{jk}$. However, in most cases
it proves simpler to work in terms of the direction cosine $\mu_{jk} \equiv 
\cos \theta_{jk}$.}
\label{geom}
\end{figure}

These approximations allow us to write the radiative transfer equation for 
an arbitrary line of sight through the protogalaxy
\begin{equation}
 \frac{\deriv{I_{\nu}}}{\deriv{s}} = - \kappa_{\nu}(s) I_{\nu} + j_{\nu}(s),
\end{equation}
(where $I_{\nu}$ is the specific intensity, $\kappa_{\nu}$ and $j_{\nu}$ 
are the opacity and emissivity respectively, and where $s$ is the position 
along the line of sight) in the simpler form
\begin{equation}
 \frac{\deriv{I_{\nu}}}{\deriv{s}} = - \kappa_{\nu}(s) I_{\nu},
\end{equation}
which can be further simplified to
\begin{equation}
 \frac{\deriv{I_{\nu}}}{\deriv{\tau_{\nu}}} = - I_{\nu}.
\end{equation}
Solution of this equation is trivial; if the specific intensity at the edge 
of the protogalaxy is $I_{\nu}(0)$, then at a distance $s$ into the 
protogalaxy, it is
\begin{equation}
 I_{\nu} = I_{\nu}(0) e^{-\tau_{\nu}(s)}.
\end{equation}
To obtain the mean specific intensity, we simply integrate this 
expression over solid angle. Using the conventional variables $r$ and $\mu$
(see figure~\ref{geom}), we find that 
\begin{equation}
 \label{meanf}
 J_{\nu}(r) = \frac{1}{2} J_{\nu}(r_{t}) \int_{-1}^{+1} \exp 
\left[-\tau(\nu,r,\mu) \right] \, \deriv{\mu},
\end{equation}
where $J_{\nu}(r_{t})$ is the incident intensity at the edge of the 
protogalaxy (i.e.\ the background radiation) and where $\tau(\nu,r,\mu)$ 
is the optical depth between $r$ and the edge of the protogalaxy in the 
direction of $\mu$.

\subsubsection{Photoionization}
\label{pi}
For the photoionization reactions (numbers 22--24, 26--27 \& 29), 
the opacity is dominated by continuum absorption by neutral hydrogen and 
helium. For static gas, we can write this as
\begin{eqnarray}
 \tau(\nu,r,\mu) & = & \sigma_{\mH}(\nu) N_{\mH}(r, \mu) 
+ \sigma_{\He} N_{\He}(r, \mu) \\
 & = &  \left[ \sigma_{\mH} + \left( \frac{Y}{4-4Y} \right) 
\sigma_{\He} \right] N_{\mH}(r, \mu),
\end{eqnarray}
where the second line follows if we assume primordial abundances and a small
fractional ionization (which also allows us to neglect $\Hep$ absorption).

Substituting this expression into equation~\ref{meanf}, we obtain
\begin{equation}
 \label{jion}
  J_{\nu}(r) = \frac{1}{2} J_{\nu}(r_{t}) \exp \left[ - \left( \sigma_{\mH} 
 + \frac{Y}{4-4Y} \sigma_{\He} \right) \right]  {\cal N}_{\mH}(r),
\end{equation}
where
\begin{equation}
 {\cal N}_{\mH}(r) = \int_{-1}^{+1} \exp \left[-N_{\mH}(r,\mu) \right] \, 
\deriv{\mu}.
\end{equation}
Aside from ${\cal N}_{\mH}(r)$, all of the terms in equation~\ref{jion} 
are known at
the beginning of the simulation. Moreover, if we assume that $J_{\nu}(r_{t})$
remains constant for the duration of the simulation, then it becomes possible
to calculate the photoionization rates in advance, tabulating them as 
functions of ${\cal N}_{\mH}$. If we do this, then instead of solving directly 
for the set of photoionization rates at each timestep, we can simply 
calculate ${\cal N}_{\mH}(r)$ and then interpolate the rates from the 
appropriate tables; this avoids a large number of frequency integrations, 
and results in a significant speed-up.

\subsubsection{Photodissociation}
\label{pd}
For Lyman-Werner photons, the situation is a little more complicated, due to 
the large number of spectral lines involved. Fortunately, there is a simple
approach that we can use which significantly reduces the amount of 
computation required to obtain the photodissociation rate.

We begin by writing the total $\mHt$ photodissociation rate as a sum of the
contributions from each individual rotational and vibrational level:
\begin{equation}
 k_{\rm dis} n_{\mHt} \equiv \sum_{j} k_{j} n_{j},
\end{equation}
where $n_{\mHt}$ is the total number density of $\mHt$ molecules, 
and $n_{j}$ is the 
number density of molecules in the $j$-th ro-vibrational level. For each level,
 we can write the photodissociation rate in the form \citep{db} 
\begin{equation}
 \label{dis_by_W}
 k_{j} = \frac{2 \pi J_{\nu}(r_{t})}{h} \int_{-1}^{+1}
 \sum_{i} f_{{\rm dis},i} \frac{\deriv{W_{ij}}}{\deriv{N_{\mHt, j}}} 
\, \deriv{\mu},
\end{equation} 
where $W_{ij}$ is the dimensionless equivalent width of the $i$-th 
Lyman-Werner line originating from level $j$, $N_{\mHt, j}$ 
is the column density of $\mHt$ molecules in level $j$, and $f_{{\rm dis},i}$
is the dissociation fraction for level $i$ (i.e.\ the fraction of excitations
that are followed by dissociation, rather than de-excitation to some other 
bound state).
 
The equivalent width can be written as
\begin{equation}
 \label{eqw}
 W_{ij}  = \int_{0}^{\infty} \left[ 1 - \exp (-\sigma_{i}(\nu) N_{\mHt, j})
\right] \frac{\deriv{\nu}}{\nu},
\end{equation}
where the radiative cross-section $\sigma_{i}(\nu)$ can be
written in terms of the line profile $\phi(\nu, \nu_{i})$ 
and oscillator strength $f_{{\rm osc},i}$ as
\begin{equation}
\sigma_{i}(\nu) = \frac{\pi e^{2}}{m_{e} c^{2}} \: f_{{\rm osc},i} 
\: \phi(\nu, \nu_{i}).
\end{equation}
In general, the form of the line profile will depend upon the details of the
temperature and velocity variations along the line of sight. For static, 
isothermal gas, however, it reduces to the familiar Voigt profile. 
Consequently, $\sigma_{i}(\nu)$ is known {\em a priori}, allowing us to 
tabulate $W_{ij}$, and hence $\deriv{W_{ij}}/\deriv{N_{\mHt, j}}$, as
functions of $N_{\mHt,j}$ at the beginning of the simulation. Similarly,
we can tabulate the photodissociation rates $k_{j}$ as a function of
\begin{equation}
 {\cal N}_{\mHt, j}(r) = \int_{-1}^{+1} \sum_{i}  f_{{\rm dis},i}
\frac{\deriv{W_{ij}}}{\deriv{N_{\mHt, j}}} \, \deriv{\mu},
\end{equation} 
since the dissociation fractions $f_{{\rm dis},i}$ are determined purely by
atomic physics and are independant of the physical conditions in the gas.
In this way, we reduce the problem of determining the photodissociation rates
to the simpler one of determining the column density of $\mHt$ molecules 
in each ro-vibrational level for a large number of lines of sight. Given
a set of column densities for a particular level, we can obtain the 
corresponding values of $\deriv{W_{ij}}/\deriv{N_{\mHt, j}}$ by 
interpolation; from these, we can derive $ {\cal N}_{\mHt, j}(r)$ by 
numerical integration; and finally, the photodissociation rate $k_{j}$ 
is obtained by another interpolation.

Two potential problems remain with this method, however. Firstly, 
equation~\ref{eqw} breaks down at high column densities as two effects that 
we have neglected -- absorption by neutral hydrogen, and overlap of the 
individual Lyman-Werner lines -- become significant. However, these 
effects are important only for $\mHt$ column densities 
$N_{\mHt} > 10^{20} \: \rm{cm}^{-2}$; since the largest $\mHt$ 
column densities that we encounter in our simulations are typically no larger
than $N_{\mHt} = 10^{18} \: \rm{cm}^{-2}$, it is generally safe to
ignore these complications.~\footnote{In any case, it is possible to 
reformulate equation~\ref{eqw} so as to avoid these difficulties,
as in \citet{db}.}

The second and more significant problem is the fact that, in the method as 
outlined, we have to deal with all of the rotational and vibrational 
levels of $\mHt$. This makes calculating the photodissociation rate much
more costly than calculating the photoionization rates, as we need to 
compute several hundred different column densities for each line of sight. 
Similarly, the chemical network is also greatly enlarged and consequently 
takes longer to solve. Moreover, the excited vibrational levels typically have 
characteristic timescales of the order of $10^{6} \: \rm{s}$, much smaller 
than the timescale on which the total $\mHt$ abundance changes, forcing us to 
solve the chemical network with a timestep that is orders of magnitude 
smaller than we could otherwise use.

To overcome these problems, we make another approximation. Since, at the 
temperatures and densities encountered in our simulations, most of the 
excited levels have very small populations, their net contribution to 
the photodissociation rate is also very small. We can therefore safely
ignore them, and concentrate only on the few rotational levels that 
are responsible for the bulk of the photodissociation.

The simplest approximation that we can make is to assume that all of the 
$\mHt$ molecules are in the form of para-hydrogen and are to be found in the 
ground state. This approximation was suggested by \citet{aanz}, and supposes 
that any ortho-hydrogen formed will be rapidly converted to para-hydrogen by
collisions with protons \citep{ger}
\begin{equation}
 \mHt({\rm ortho}) + \Hp \rightarrow \mHt({\rm para}) + \Hp.
\end{equation}
Unfortunately, this is only valid at very low temperatures, typically 
$T \simless 50 \: \rm{K}$. At higher temperatures, conversion from 
para-hydrogen back to ortho-hydrogen via the inverse reaction becomes 
competitive. Moreover, at realistic temperatures and densities, ortho-para 
interconversion via proton collisions with excited states will also occur,
and in general the ortho to para ratio will be close to its equilibrium
value of three to one.

We can therefore obtain a much better approximation to the true 
photodissociation rate by including ortho-hydrogen. For simplicity, we 
assume that the ortho to para ratio is fixed and is exactly three to one.
We continue to ignore the other excited rotational levels.

Comparing the results of simulations performed using this approximation
with those of simulations in which the rotational and vibrational level 
populations are set to their local thermodynamic equilibrium (LTE) values,
we find that it performs well: although differences in the final $\mHt$ 
abundances are often observed, the effect on $T_{\rm crit}$ remains small
-- it is increased by at most 30--40\% at any given redshift, and by much 
less than this in the limits of low and high redshift. The uncertainty that
this introduces is comparable to that due to our poor knowledge of the
high-redshift ultraviolet background (see figure~\ref{baseline} below).
Moreover, since the true level populations will lie somewhere between our 
simple approximation and the LTE case, this suggests that our approximation 
is sufficient for our current purposes, and that a more detailed 
(and hence slower) treatment is not required.  

\subsubsection{Calculating the column densities}
As demonstrated in the previous sections, we can determine the photochemical 
reaction rates for a given mass shell by computing the functions 
\begin{equation}
\label{cal1}
 {\cal N}_{\mH}(r) = \int_{-1}^{+1} \exp \left[-N_{\mH}(r,\mu) \right] \, 
\deriv{\mu},
\end{equation}
and 
\begin{equation}
\label{cal2}
 {\cal N}_{\mHt, j}(r) = \int_{-1}^{+1} \sum_{i} f_{{\rm dis},i} 
\frac{\deriv{W_{ij}}}{\deriv{N_{\mHt, j}}} \, \deriv{\mu},
\end{equation}
(where $j=0,1$ for the para and ortho-hydrogen ground states 
respectively) and then interpolating the rates from a pre-built look-up table.
Moreover, knowledge of $N_{\mH}(r,\mu)$ and $N_{\mHt, j}(r,\mu)$ 
is sufficient to specify ${\cal N}_{\mH}(r)$ and ${\cal N}_{\mHt, j}(r)$
respectively, so the problem is essentially reduced to one of calculating
these column densities.

To do this, we use a method very similar to that in
\citet{kbs}. For each mass shell $k$, we choose a point $r_{k}$ at the centre
of the shell (see figure~\ref{geom}). For each of these points, we compute
the column densities along $N_{a}$ lines of sight with uniform angular 
separation. The column density of a chemical species $i$ along one of these 
lines of sight is given by  
\begin{equation}
 \label{cd}
 N_{i}(r,\mu) = \int_{0}^{L} n_{i}[r'(x)] \, \deriv{x},
\end{equation}
where $L= r\mu + [r_{t}^{2} - r^{2}(1-\mu^{2})]^{1/2}$ 
is the total distance along the ray to the edge 
of the protogalaxy, and where $r'(x) = (r^{2}+x^{2}-2rx\mu)^{1/2}$ is the 
distance from the point labeled by $x$ to the centre of the protogalaxy.
Solution of this equation by means of numerical integration is straightforward.

Given the set of column densities for a particular point $r_{k}$,
we use equations~\ref{cal1} and \ref{cal2} to compute ${\cal N}_{\mH}(r_{k})$  
and ${\cal N}_{\mHt, j}(r_{k})$,~\footnote{In the latter case, we make use
of molecular data from \citet{ably,abwer} and \citet{rou}, together with 
dissociation fractions from \citet{ard}.} and then finally use these values to 
compute the photochemical rates at that point. Finally, the spherical 
symmetry of the problem allows us to generalize these rates to the whole
of the mass shell containing $r_{k}$. 

The accuracy of this procedure clearly depends both upon the spatial resolution
of the grid ($N_{g}$) and upon the angular resolution with which we sample it
($N_{a}$). However, as the density, temperature and chemical abundances all
vary smoothly within the protogalaxy, we find that very high resolution is
not required: setting $N_{g}=100$ and $N_{a}=20$ proves to be sufficient for 
accurate calculation of the photochemical rates. Simulations run with 
substantially higher values produce very similar results.

\subsection{Heating and cooling}
\label{cool}
The heating and cooling processes included in our model are listed in 
table~\ref{cooltab}, together with references to the sources for the rates
adopted. At low temperatures, cooling is dominated by rotational and 
vibrational line emission from $\mHt$; we include this according to the 
recent prescription of \citet{lpf}, which is probably the most accurate
available to date. We do not include the effects of HD cooling: this
becomes significant only at very low temperatures, and in any case does
not appear to have a great effect on the evolution of the gas \citep{bcl2}. 

At high temperatures ($T \simgreat 8000 \: {\rm K}$), cooling through the 
atomic lines of hydrogen (often referred to simply as Lyman-$\alpha$ cooling) 
rapidly becomes dominant. This sets an upper limit on $T_{\rm crit}$; 
for simplicity, we neglect any redshift dependence of this limit, and assume
that protogalaxies with $T_{\rm vir} \geq 8000 \: \rm{K}$ can always cool 
successfully.

Heating of the protogalactic gas is driven primarily by the photoionization
of hydrogen and helium. $\mHt$ photodissociation also contributes to the 
heating rate, but is generally not significant, due to the small $\mHt$ 
abundance.

\begin{table}
\begin{tabular}{cc}
\hline
\multicolumn{1}{c}{Process} & \multicolumn{1}{c}{Reference} \\
\hline
Atomic line cooling: & \\ 
 \hi & \citet{cen} \\
 \hei & \citet{bbft} \\
 \heii & \citet{acku} \\
 & \\
Molecular line cooling: & \\
 $\mHt$ & \citet{lpf} \\
 & \\
Collisional ionization: & \\
 \hi & \citet{janev} \\
 \hei & \citet{janev} \\
 \heii & \citet{aanz} \\
 & \\
Recombination: & \\
 \hii & \citet{fph} \\
 \heii & \citet{ap} \\
 \heiii & \citet{fph} \\
 & \\
Other chemistry: & \\
$\Hm$ formation & \citet{sk} \\
$\mHtp$ formation & \citet{sk} \\
 & \\
Bremsstrahlung: & \citet{spit} \\
 & \\
Compton scattering: & \citet{peeb} \\
 & \\
Photoionization: & \\
 \hi & \citet{os}, \citet{ysd} \\
 \hei & \citet{ysd} \\
 \heii & \citet{os} \\
 & \\
Photodissociation: & \\
 $\mHt$ & \citet{bd} \\
\hline
\end{tabular}
\caption{The main processes responsible for heating and cooling metal-free gas,
together with references to the source(s) from which the heating and cooling
rates used in our simulations were taken.} 
\label{cooltab}
\end{table}

\subsection{Running the simulation}
Given the initial temperature and chemical abundances, plus the set of
chemical reaction rates, actually solving for the thermal and chemical
evolution is relatively easy. As in the IGM case, we simply solve the
coupled set of chemical rate equations
\begin{equation}
 \frac{\deriv{n_{i}}}{\deriv{t}} = C_{i}(n_{j},T) - D_{i}(n_{j},T)n_{i}, 
\end{equation}
\begin{equation}
 \frac{\deriv{T}}{\deriv{t}} = \frac{2}{3 k n_{\rm tot}} 
\left( \Gamma - \Lambda \right),
\end{equation}
using the {\sc stifbs} integrator of \citet{nr}.

At the start of each timestep (hereafter time $t$), we compute the 
photochemical rates as outlined in sections~\ref{rtmethod}. We then use
{\sc stifbs} to solve for the new chemical abundances and new temperature
at the end of the timestep (time $t + \Delta t$), repeating this for each 
shell in turn. We next store these values, return to time $t$, and 
recalculate them using the same procedure, but with {\em two} timesteps of 
length $\Delta t/2$. We recalculate the photochemical rates at the 
intermediate point. 

We next test for convergence by comparing our two sets of results. If any 
of the chemical abundances or temperatures of any of the shells differ by 
more than 0.1\%, then we reject the results and begin again from  
time $t$ with a smaller timestep. Otherwise, we check to see whether we
need to halt the simulation, using the criteria discussed below, and, if we do
not, we begin the computations for a new timestep starting from $t + \Delta t$.
    
One final approximation that we find useful in practice is to fix the
$\Hm$ and $\mHtp$ abundances at their equilibrium values. This allows the
integrator to take much larger timesteps than would otherwise be possible,
but introduces very little error into the computed $\mHt$ abundances.

\subsection{Halting the simulation}
\label{halt}
Using the method outlined in the preceding sections, we compute the 
chemical and thermal evolution of the protogalaxy until one of two conditions 
is met: either the protogalactic gas begins to cool strongly, or we exceed a 
preset time limit, $t_{\rm lim}$.

To assess whether gas cooling is `strong' enough requires an objective
cooling criterion. A number of different possibilities have been suggested 
in the literature \citep{ro,teg,har}. In our simulations, we adopt the 
criterion used by \citet{har}: we require that the elapsed time exceeds the 
cooling time, as computed at the edge of the protogalactic core, at a 
distance $r_{0} = 0.034 r_{t}$ from the centre of the protogalaxy. 
The advantage of this choice is 
that it avoids giving us a false positive result in cases where 
$t_{\rm cool}$ drops briefly below $t_{\rm dyn}$ at early times, but remains 
so for a time $t \ll t_{\rm cool}$. As an additional sanity check, we also
require that the final temperature be smaller than the initial temperature.

If the protogalactic gas does not cool strongly, then the simulation will
terminate when it reaches $t_{\rm lim}$. This pre-set time limit is required on
purely practical grounds, to prevent simulations in which the gas does not
cool from running for excessive amounts of time, but also has a physical 
justification. In our simulations, we treat protogalaxies as isolated 
objects, uninfluenced by external events. In reality, they are part of
a dynamically evolving mass distribution,
and the majority will only survive for a limited time before merging 
with other protogalaxies of a similar or larger size.
It is possible to use the Press-Schechter formalism to calculate the 
distribution of survival times as a function of mass \citep{laco}, but for
rare objects the mean survival time is typically of the order of the Hubble 
time and thus for simplicity we set $t_{\rm lim} = t_{\rm H}$. 

\section{Results}
\label{res}
In the following sections, we present results from a number of simulations
that examine the effects of the X-ray backgrounds produced by the various
source models discussed in sections~\ref{def_xray} and \ref{alt_xray}.
Unless otherwise noted, all of these simulations assume the same parameters
for the star formation model: a standard Salpeter IMF, with 
$M_{\rm min} = 0.1 \: \rm{M}_{\odot}$ and $M_{\rm max} = 100 \: 
\rm{M}_{\odot}$,
a star formation efficiency $\epsilon = 0.1$ and a starburst lifetime 
$t_{\rm on} = 10^{7} \: \rm{yr}$. Additionally, all of the simulations
use the same cosmological model, the $\Lambda$CDM concordance model of 
\citet{wang}, which has parameters $(\Omega_{\Lambda}, \Omega_{m}, 
\Omega_{b}, h, n, \sigma_{8}) = (0.67, 0.33, 0.041, 0.65, 1.0, 0.9)$.  

\subsection{Simulations without an X-ray background}
\label{res_nox} 
Before examining the effect of an X-ray background on protogalactic
evolution, we first briefly study the evolution of $T_{\rm crit}$ 
in its absence. As well as allowing us to determine the sensitivity of 
our results to variations in the UV source model, this also provides us with a
necessary baseline against which to compare our other results. 

\begin{figure}
\centering
\epsfig{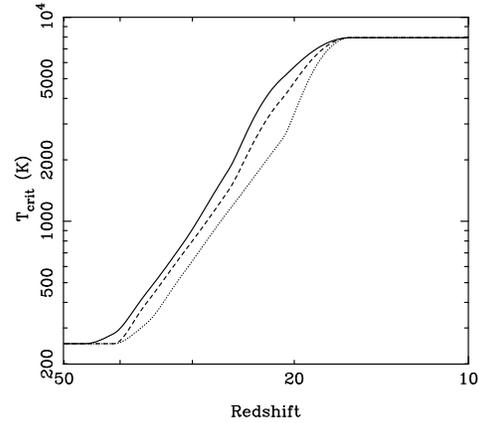}
\caption{The evolution of the critical temperature as a function of redshift
in the absence of an X-ray background. Results are plotted for three
different UV source models: two metal-free models, one with a minimum stellar 
mass $M_{\rm min} = 0.1 \: \rm{M}_{\odot}$ and the other with 
$M_{\rm min} = 1 \: \rm{M}_{\odot}$ (represented by the dotted and dashed lines
respectively) and a $Z = 0.02 Z_{\odot}$ model with 
$M_{\rm min} = 1 \: \rm{M}_{\odot}$ (solid line). In all three cases, the IMF 
is of Salpeter form, with a maximum mass $M_{\rm max} = 100 \: \rm{M}_{\odot}$.
The remaining parameters of the star formation model (star formation 
efficiency and starburst lifetime) are the same in all three 
simulations, as is the cosmological model.}
\label{baseline}
\end{figure}

In figure~\ref{baseline}, we plot the evolution of $T_{\rm crit}$ with 
redshift for the three different star formation models discussed in 
section~\ref{uvdensity}. All three models assume a Salpeter IMF,
with maximum stellar mass $M_{\rm max} = 100 \: \rm{M}_{\odot}$, 
as well as our standard star formation efficiency, starburst lifetime 
and cosmological model, described previously. Our basic model assumes
a metal-free stellar population, with a minimum stellar mass of 
$M_{\rm min} = 0.1 \: \rm{M}_{\odot}$; the corresponding results are 
given by the dotted line in figure~\ref{baseline}. The dashed line 
illustrates the effect of increasing the minimum mass to 
$M_{\rm min} = 1 \: \rm{M}_{\odot}$; the solid line assumes the same 
$M_{\rm min}$, together with a metallicity $Z = 0.02 Z_{\odot}$.

Figure~\ref{baseline} demonstrates that although the strength of the 
Lyman-Werner background increases by almost a factor of five as we move
from our basic model to the metal-enriched model, this has less effect on
$T_{\rm crit}$ than we might expect: the difference in $T_{\rm crit}$ between
the three models is never more than 50\%, and the qualitative details
of its evolution are very similar in all three models. This suggests 
that the uncertainty introduced by our lack of knowledge of the 
properties of the primordial stellar population need not be unduly limiting. 
However, it is also clear that some uncertainty remains, and this will place
a lower limit on the magnitude of any effect that we can reliably claim to 
detect, as small variations in $T_{\rm crit}$ due to the X-ray background 
will be swamped the by the error resulting from the uncertainty in the 
Lyman-Werner background.

In the work that follows, we take as our baseline the results of our basic,
metal-free, low $M_{\rm min}$ model; as figure~\ref{baseline} demonstrates,
this minimizes the strength of the Lyman-Werner background, and thus will 
tend to maximize the effectiveness of the X-ray background.

\subsection{Massive X-ray binaries}
\begin{figure}
\centering
\epsfig{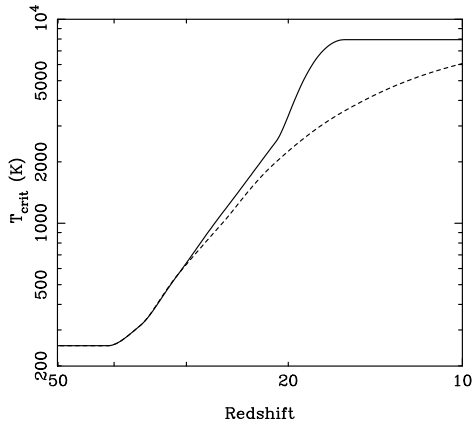}
\caption{The evolution of $T_{\rm crit}$ as a function of redshift
 for an X-ray background produced by the massive X-ray binary model 
 of section~\ref{def_xray} (dashed line). For comparison, we also plot 
 the evolution of $T_{\rm crit}$ for our basic X-ray free model (solid line).}
\label{base_xray}
\end{figure}

In figure~\ref{base_xray}, we plot the evolution of $T_{\rm crit}$ 
in the presence of the X-ray background generated by the massive X-ray 
binary model described in section~\ref{def_xray}. For the purposes of 
comparison, we also plot the results of our basic X-ray free model. 
Initially, the evolution of $T_{\rm crit}$ is the same in both models,
implying that the X-ray background has little or no effect on the gas. 
At a redshift $z \simeq 20$ and critical temperature $T_{\rm crit} \simeq 
2000 \: \rm{K}$, however, the models begin to diverge significantly. 
In the X-ray free model, the critical temperature continues to increase 
rapidly until it reaches its maximum value of $T_{\rm crit} = 8000 \: \rm{K}$. 
In the X-ray binary model, by contrast, the rate of increase of $T_{\rm crit}$ 
is significantly slower, and it fails to reach its maximum value by the 
end of the simulation at $z=10$.~\footnote{We choose to end our simulations
at $z = 10$ because at lower redshift we expect the effects of ionizing
radiation from both stellar sources and quasars to become increasingly 
important and thus our results to become unreliable. Clearly if
reionization occurs at $z > 10$, the same is true for some portion of the
results plotted here.}

It is clear from figure~\ref{base_xray} that the presence of the X-ray
background significantly affects the evolution of gas in the larger of 
the $\mHt$-cooled protogalaxies. In small protogalaxies, on the other hand,
the negative feedback caused by the Lyman-Werner background remains as strong
as ever. 

A simple way to judge the importance of this effect is to examine the
difference it makes to the fraction of gas in the universe that 
can collapse and cool. As we saw in section~\ref{sfr}, we can use the 
Press-Schechter formalism to express the cool gas fraction as
\begin{equation}
 F(z,M_{\rm crit}) = \rm{erfc}\left[ \frac{\delta_{c}(z)}{\sqrt{2} 
\sigma(M_{\rm crit})} \right]. 
\end{equation}
Using the relationship between $M_{\rm crit}$ and $T_{\rm crit}$ derived in 
that section, it is straightforward to calculate the evolution of 
$F(z,M_{\rm crit})$ in both the X-ray binary and X-ray free models.

\begin{figure}
\centering
\epsfig{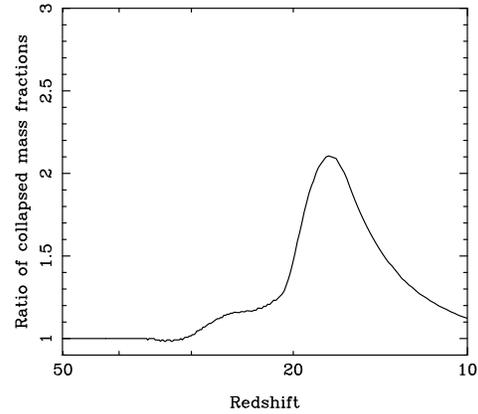}
\caption{The ratio of the cooled gas fraction in the massive X-ray binary 
model to that in the X-ray free model. The mass fractions are calculated 
using the Press-Schechter formalism, as outlined in the text.}
\label{fcoll}
\end{figure}

To better highlight the difference between the two models, 
we plot in figure~\ref{fcoll} the ratio of the cooled gas fraction in
the X-ray binary model to that in the X-ray free model. At high redshift, the 
evolution of $T_{\rm crit}$ is the same in both models, and thus the ratio is
one. Below $z = 30$, the behaviour of the models begins to diverge, but this
has little effect on the ratio until the models begin to diverge sharply
at $z \sim 20$. Thereafter, the ratio rises sharply to a peak at $z \sim 15$,
and subsequently declines as the growth in the cooled mass fraction becomes 
dominated by the formation of larger protogalaxies that cool via 
Lyman-$\alpha$ emission.

Figure~\ref{fcoll} shows us that by ignoring the effect of the X-ray 
background we underestimate $F(z,M_{\rm crit})$ by at most a factor of 
two. In fact, the difference is likely to be even smaller: we have assumed
that all of the gas in a protogalaxy with $M > M_{\rm crit}$ will cool, but
this need not be the case if an X-ray background is present, as X-ray heating
will prevent gas from cooling in the low-density outer layers of the 
protogalaxy. Whether this will affect the star formation rate within individual
protogalaxies is not clear, but in any case the difference in the 
globally-averaged star formation rate is unlikely to be greater than a factor
of two. Note, however, that this extra star formation occurs entirely in 
low-mass systems, from which ionizing photons \citep{rs} and supernova ejecta 
\citep{ft} can more readily escape. Doubling the star formation rate may thus
significantly increase the feedback of primordial star formation on the IGM.

\begin{figure}
\centering
\epsfig{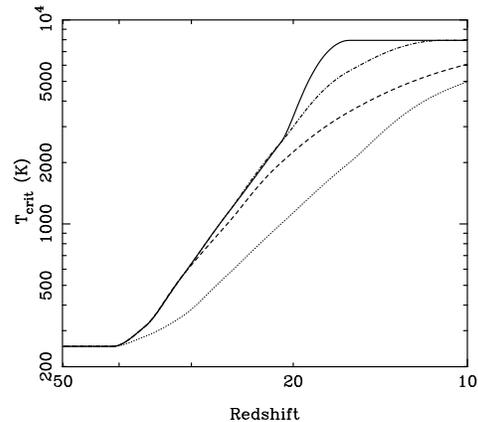}
\caption{As figure~\ref{base_xray}, but including variants of the 
 X-ray binary model with 10\% (dash-dotted line) and 1000\% (dotted line) 
 of the flux of the basic model (dashed line). As before, we also plot
 the evolution of $T_{\rm crit}$ for the X-ray free model (solid line).}
\label{alt_norm}
\end{figure}

Finally, given the uncertainties in the data underlying our simple massive 
X-ray binary model (see section~\ref{def_xray}), it is of interest to
investigate the sensitivity of our results to changes in the strength of
the X-ray background. Accordingly, we have run additional simulations in 
which the strength of the X-ray sources was increased or decreased by a 
factor of ten. The results are plotted in figure~\ref{alt_norm}, together
with the results of our basic X-ray binary model and of the X-ray free 
model.

Unsurprisingly, increasing or decreasing the strength of the X-ray 
background alters the evolution of $T_{\rm crit}$. Decreasing it by an
order of magnitude increases $T_{\rm crit}$ to the point where its
evolution is little different from that in the X-ray free model.
Increasing it by an order of magnitude, on the other hand, 
systematically lowers $T_{\rm crit}$, although never by more than 
a factor of two.

\subsection{Supernova remnants} 
As we discussed in section~\ref{alt_xray}, it is possible that massive X-ray
binaries are much less abundant at high redshift than at the present day.
If so, then the X-ray emission of star-forming galaxies will be dominated
by supernova remnants (SNR). These will generate X-rays through two 
main emission mechanisms: thermal bremsstrahlung from hot gas, and inverse 
Compton scattering of the CMB by relativistic electrons. We consider the 
effects of these mechanisms separately.

We examined two possible variants of the thermal bremsstrahlung model. In one, 
we assumed that the characteristics of the emission are broadly the same 
as those observed locally, with a fraction $f_{\rm x} = 2 \times 10^{-4}$ 
of the supernova energy being radiated as X-rays with a characteristic 
temperature
$T_{\rm x} = 1 \: \rm{keV}$ \citep{helf}. In the other model, we assumed that
all high-redshift supernovae explode in extremely dense surroundings,
producing X-ray bright, ultra-compact remnants with $f_{\rm x} = 0.01$ and
$T_{\rm x} = 30 \: \rm{keV}$ \citep{csnr}. Realistic models should lie
somewhere between these two extremes. 

However, we found that in neither of these cases does the X-ray background 
affect the 
evolution of $T_{\rm crit}$: at our level of temperature resolution, the 
results are identical to those obtained for the X-ray free model. The obvious
conclusion is that the background produced by bremsstrahlung is simply too 
weak to be effective.

\begin{figure}
\centering
\epsfig{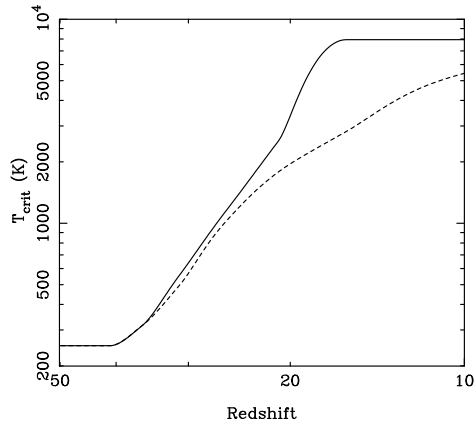}
\caption{As figure~\ref{base_xray}, but for an X-ray background produced by
inverse Compton emission from supernova remnants. We assume that 10\%
of the total supernova energy is transferred to the relativistic electrons
powering the emission, and again plot the results of the X-ray free model
for comparison.}
\label{SN_IC}
\end{figure}

Our inverse Compton model fares somewhat better. In this model, the strength
of the X-ray background is proportional to the mean fraction $f_{\rm e}$ 
of the supernova explosion energy that is transferred to relativistic electrons
within the remnant. This value is not known accurately and so we considered 
two possible cases, one with $f_{\rm e} = 10^{-3}$ and another with 
$f_{\rm e} = 0.1$, these being conservative lower and upper bounds on the 
true value.
In the former case, we again saw no significant effect on the evolution of 
$T_{\rm crit}$. In the latter case, on the other hand, we saw results very
similar to those obtained for the X-ray binary model, as illustrated in 
figure~\ref{SN_IC}.

In principle, therefore, inverse Compton emission 
from supernova remnants could be as important an X-ray source as emission
from X-ray binaries. Ultimately, however, its importance depends upon the 
value of $f_{\rm e}$, and no firm conclusions are possible until this
value is better constrained. In this context, the possible observational 
tests suggested by \citet{oh} could prove extremely valuable.

\subsection{The thermal and chemical evolution of the IGM}
\label{ther_chem}
\begin{figure}
\centering
\epsfig{figure=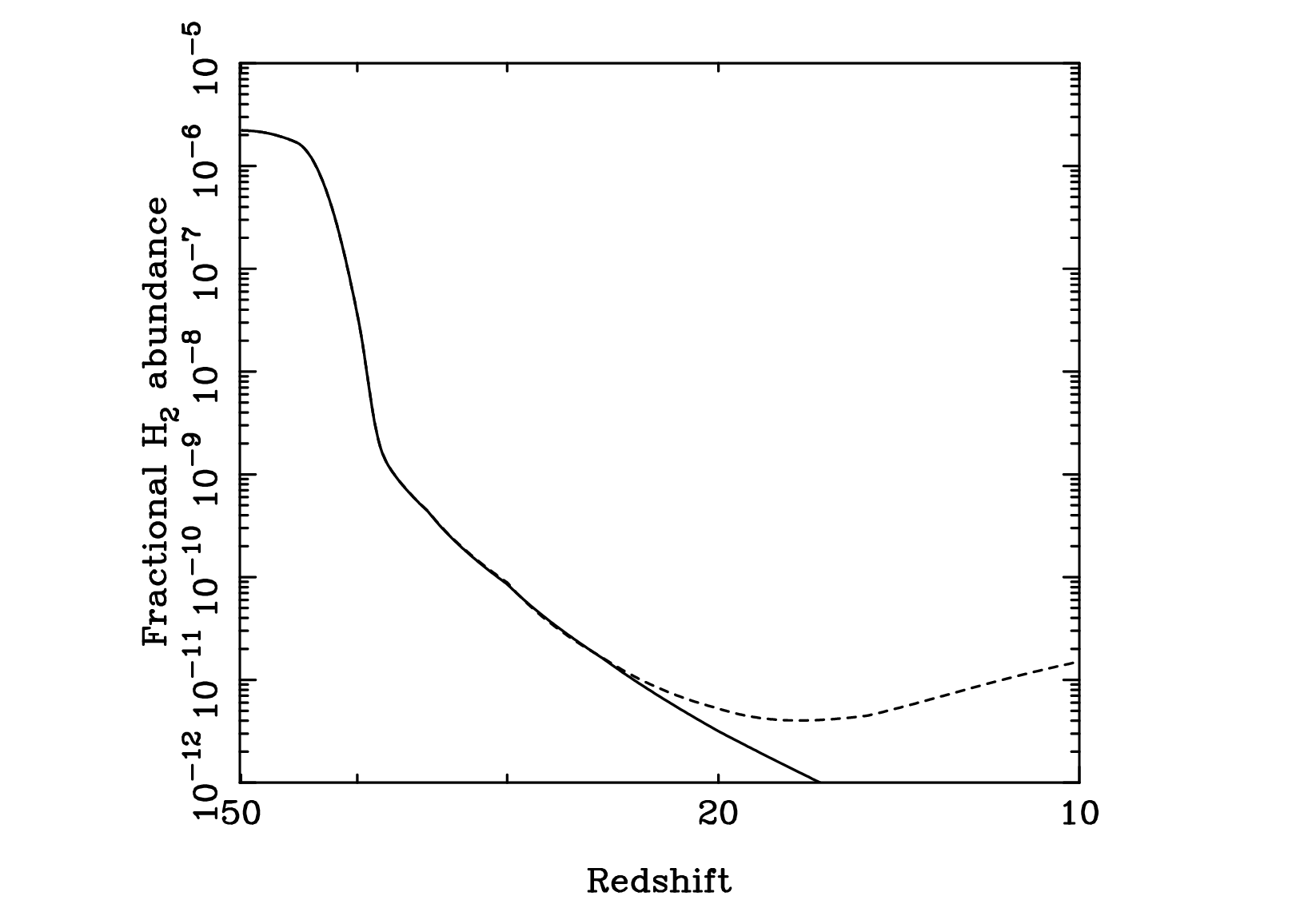,width=5.8cm,angle=270,clip=}
\caption{The evolution with redshift of the fractional abundance of $\mHt$ 
in the intergalactic medium, plotted for the X-ray free model (solid line)
and the massive X-ray binary model (dashed line).}
\label{h2_in_igm}
\end{figure}

As well as determining the evolution of $T_{\rm crit}$, our simulations 
also allow us to study the thermal and chemical evolution of the diffuse 
intergalactic medium, as outlined in section~\ref{model_pg}.
As a simple example, we plot in figure~\ref{h2_in_igm} the evolution 
of the fractional $\mHt$ abundance in the IGM for the X-ray free 
model (solid line) and the X-ray binary model (dashed line).

In both cases, the fractional abundance rapidly decreases from its primordial
value due to photodissociation by the ultraviolet background, reaching 
$f_{\mHt} = 10^{-9}$ by $z \simeq 35$. Subsequently, its decline slows, in 
part because the rate of increase in the strength of the ultraviolet 
background also slows. Below $z = 20$, the behaviour of the two models 
diverges. In the X-ray free model, $f_{\mHt}$ continues to decline until the 
end of the simulation. In the X-ray binary model, on the other hand, the 
increasing ionization of the IGM boosts the $\mHt$ formation rate to the 
point where it overtakes the photodissociation rate and the $\mHt$ abundance,
after reaching a minimum at $z \simeq 17$, begins to climb. 
Nevertheless, it remains
extremely small at the end of the simulation, readily justifying our 
assertion in section~\ref{lwabs} that absorption by intergalactic $\mHt$ 
does not play a significant role in determining the strength of the 
Lyman-Werner background.  

\begin{figure}
\centering
\epsfig{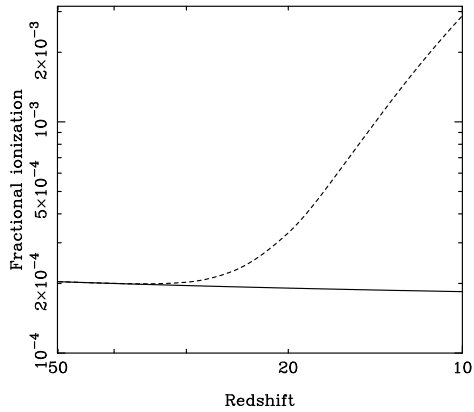}
\caption{The evolution with redshift of the fractional ionization of the 
intergalactic medium, plotted for the X-ray free model (solid line)
and the massive X-ray binary model (dashed line).}
\label{igm_ion}
\end{figure}

In figure~\ref{igm_ion}, we plot the evolution
of the fractional ionization of the IGM. In the X-ray free model, this 
remains approximately constant over the lifetime of the simulation,
as the recombination timescale is significantly longer than the Hubble time.
In the X-ray binary model, on the other hand, photoionization by the growing 
X-ray background eventually overcomes the very small recombination rate and  
drives the fractional ionization upwards, increasing it by just over an order
of magnitude by the end of the simulation. Even so, it remains small, 
demonstrating that the X-ray background does not contribute significantly
to cosmological reionization.

\begin{figure}
\centering
\epsfig{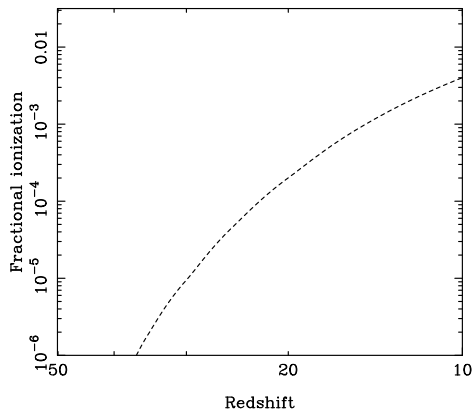}
\caption{The evolution with redshift of the fractional ionization of 
helium in the intergalactic medium, plotted for the massive X-ray binary 
model. In the X-ray free model, the $\Hep$ abundance remains negligible 
throughout the simulation.}
\label{igm_heion}
\end{figure}

The effect on the fractional ionization of helium (the ratio
of $\Hep$ to $\He$) is rather more striking. The post-recombination 
$\Hep$ abundance is extremely small \citep{sld}, and in the X-ray free model 
remains at this low level throughout the simulation. In the X-ray binary 
model, on the other hand, it increases dramatically over the course of the 
simulation, reaching $f_{\Hep} = 4 \times 10^{-3}$ by $z = 10$. This is
illustrated in figure~\ref{igm_heion}.

\begin{figure}
\centering
\epsfig{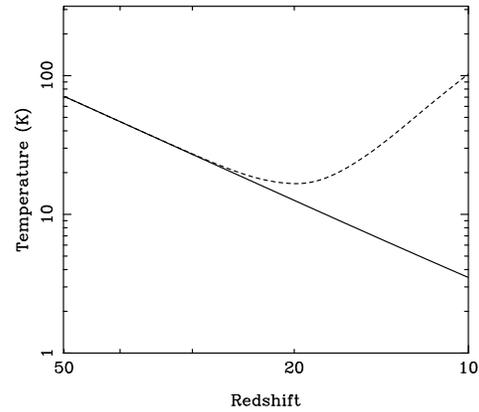}
\caption{The evolution with redshift of the temperature of the 
intergalactic medium, plotted for the X-ray free model (solid line)
and the massive X-ray binary model (dashed line).}
\label{igm_heat}
\end{figure}

Finally, in figure~\ref{igm_heat} we plot the evolution with redshift of
the temperature of the IGM. In the X-ray free model, adiabatic cooling 
dominates the thermal evolution, and the temperature falls off 
approximately as $(1+z)^{2}$. In the X-ray binary model, on the
other hand, photo-electric heating begins to heat the IGM strongly at
$z=20$, driving the temperature up to $T = 100 \: \rm{K}$ by the end of
the simulation.

Thus, although the X-ray background does not contribute significantly
to the reionization of the IGM, it does produce substantial
reheating prior to reionization. Moreover, given the large mean free path 
of the X-ray photons, this reheating occurs almost uniformly throughout the 
IGM, rather than being localized to the vicinity of star-forming galaxies.

One consequence of this reheating is that the formation of very small-scale 
structure will be suppressed, as the increased temperature of the IGM leads 
to an increased Jeans mass. This is unlikely to affect the global star 
formation rate, however, as star formation within these small structures 
would in any case be strongly suppressed by the ultraviolet background.
Nevertheless, it will reduce the mean clumping factor of the IGM below the
level that we would otherwise predict, which may in turn speed up 
reionization \citep{ham,blob}.

Reheating also affects the visibility of the IGM in the redshifted
$21 \: \rm{cm}$ line of neutral hydrogen. \citet{mmr} show that scattered 
Lyman-$\alpha$ emission from high redshift galaxies efficiently couples 
the spin temperature of the \hi hyperfine levels to the kinetic temperature of 
the gas. If the gas temperature is smaller than the CMB temperature, this 
results in $21 \: \rm{cm}$ line absorption; if it is greater, then it
results in emission. Absorption is easier to detect than emission \citep{sr},
but the heating produced by the X-ray background implies that absorption
occurs only at $ z \geq 15 $, and that concentrating on detecting 
$21 \: \rm{cm}$ emission may be the more viable strategy.

\section{Conclusions}
\label{conc}
The results of the previous section allow us to assess the impact
of the high redshift X-ray background that is produced by 
star-forming galaxies. If we assume that the X-ray emission of these 
galaxies is similar to that 
observed locally, and that the same correlation between X-ray luminosity and
star formation rate applies, then we find that the background produced is 
strong enough to partially offset the effects of UV photodissociation in 
large ($T_{\rm vir} > 1000 \: \rm{K}$), $\mHt$-cooled protogalaxies.

However, local emission is dominated by massive X-ray binaries, which 
may not form in large numbers at high redshift. Therefore, we have also
explored the effect of an X-ray background produced by emission from 
supernova remnants. If this emission is dominated by inverse Compton 
scattering and if the fraction of the supernova energy transferred to the 
relativistic electrons powering this emission is large, then the 
resulting background has very similar effects to one produced by X-ray 
binaries. On the other hand, if the fraction of energy transferred is small,
then the background has little or no effect. 

In addition to inverse Compton emission, we have also examined the effect
of thermal bremsstrahlung emission from hot gas in the remnants, and find
that even if all supernovae were to form X-ray bright, ultra-compact remnants,
the resulting X-ray background would still be too weak to significantly affect
protogalactic evolution.

Finally, none of these models produces an X-ray background that is strong
enough to balance UV photodissociation in small protogalaxies, with virial 
temperatures $T_{\rm vir} < 1000 \: \rm{K}$. In these protogalaxies, 
negative feedback always dominates. 

How significant are these results? One simple way to assess this is to study
the evolution of the mass fraction of cooled gas, which represents the total
amount of matter available to form stars. Comparing its evolution in the
X-ray binary model to that in the absence of an X-ray background, we find 
that it is increased by approximately a factor of two. Given our star 
formation model, this corresponds to an increase in the global star formation 
rate by the same amount. However, this is small compared to the order of
magnitude increase that would result if we were simply to ignore the effect 
of the UV background \citep[see figure 7 of][]{har}.

In reality, the difference between the two models may be greater than this
because the additional star formation takes place entirely in low-mass
systems, from which ionizing photons and supernova-produced metals can 
readily escape \citep{rs,ft}. Doubling the star formation rate may therefore
have more than double the impact on the intergalactic medium. However,
to properly assess the ultimate importance of this effect requires more
detailed modelling, beyond the scope of this paper.

Ultimately, understanding the history of star formation in the
small protogalaxies studied in this paper remains important even if they
do not contribute to the reionization or enrichment of the IGM to any great
degree. This is simply because, in a hierarchical universe, these 
protogalaxies are the building blocks from which larger galaxies form
and therefore set the initial conditions for later stages of galaxy formation. 
In particular, very little metal enrichment of the primordial gas is
required in order to allow the CNO cycle to operate and population II
(rather than population III) stars to form, and yet this can have a profound 
effect on the predicted spectral energy distribution of an early stellar 
population.

Although the main purpose of our study was to examine the effects of 
the X-ray background on the thermal and chemical evolution of gas
within protogalaxies, our approach also allows us to examine the effects
of the background on the diffuse IGM. Our main results are three-fold:
\begin{enumerate}
 \item We confirm the rapid destruction of $\mHt$ in the intergalactic 
 medium noted by \citet{har}, but also show that when an X-ray background
 is present the $\mHt$ abundance does not continue to decline indefinitely,
 but eventually stabilizes and may even begin to increase. However, it never
 becomes large enough to significantly affect the Lyman-Werner background.

 \item We show that although photoionization by the X-ray background
 significantly increases the fractional ionization of the IGM (and in
 particular the fractional ionization of helium), the bulk of the gas remains 
 mostly neutral, demonstrating that the contribution of the X-ray background 
 to cosmological reionization is small.

 \item We find that the X-ray background will also heat the intergalactic
 gas, raising its temperature to $T \simeq 100 \: \rm{K}$ by $z = 10$ 
 (compared to $T = 3.5 \: \rm{K}$ in the X-ray free model). This will
 suppress the formation of structure on the smallest scales by increasing 
 the Jeans mass. It is unlikely to affect the global star 
 formation rate, since $T_{\rm crit} \gg 100 \: \rm{K}$, but may speed
 up the process of reionization by reducing the mean clumping factor of 
 the IGM.
\end{enumerate}

A number of other authors have studied the effects of radiative feedback
on the formation of $\mHt$-cooled protogalaxies 
\citep*{cfgj,cfa,hrlx,hrl97,har,mba,mba2}. Most of these
studies have concentrated solely on the effects of the Lyman-Werner 
background, generally finding that it suppresses cooling (and hence star 
formation) by $z \sim 20$--30, prior to cosmological reionization. Our X-ray
free simulations support this conclusion. In particular, comparison with the
results of \citet{har}, who use a very similar method but with a different
implementation of radiative transfer, gas chemistry and cooling, shows good
agreement.

Rather less work has been done on the effects of the high redshift X-ray 
background. \citet{hrlx} examined the formation of molecular hydrogen
in a constant density primordial cloud illuminated by a power-law UV
spectrum extending into the hard X-rays, and found that in some cases, 
the background could enhance $\mHt$ formation. This lead \citet{har} to
examine the effects of a quasar-produced X-ray background, using a very
simple model in which the X-ray flux is a fixed fraction of the 
Lyman-Werner band flux, modulated by absorption by a fixed column density of
$\mH$ and $\He$. They found that an X-ray to UV flux ratio of 10\% was 
enough to overcome negative feedback, and that a higher flux ratio could
potentially produce positive feedback. Recently, \citep{mba2} have studied
the effects of similar model backgrounds using an adaptive-mesh hydrodynamics
code. They also find that X-rays reduce the effectiveness of negative 
feedback, but that the latter still dominates. They do not find evidence for
the positive feedback predicted by \citet{har}. However, their simulations do
not include the effects of $\mHt$ self-shielding, and thus potentially 
underestimate the amount of $\mHt$ that forms \citep[although see][for a 
different view]{mba}.

There are several significant differences between our work and these previous
investigations. Firstly, we do not assume a fixed spectrum or intensity for
the X-ray background; rather, we specify the properties of the X-ray {\em
sources} and subsequently compute the build-up of the background in a 
self-consistent fashion. Moreover, we consider source models where the
X-ray emission is proportional to the star formation rate, as is observed to
be the case for star-forming galaxies at low redshift; the relationship between
star formation rate and X-ray emission in quasar-based models is far less 
clear. As a result, we generally consider X-ray backgrounds signficantly 
fainter than those studied in the papers cited above. This makes direct
comparison of our results difficult. However, we note that \citet{mba2}
find X-ray feedback to be ineffective below $M_{\rm vir} = 10^{6} \: 
{\rm M}_{\odot}$, regardless of the strength of the X-ray background; at
the redshifts they consider, this corresponds to a virial temperature
$T_{\rm vir} \sim 1000 \: \rm{K}$, and thus agrees well with the similar 
result obtained in this paper.

In closing, we note that a number of uncertainties still remain in 
our treatment of this problem. Some of these -- the high redshift star 
formation rate or the appropriate population III initial mass function, for 
instance -- we simply do not know at the present time. Fortunately,  
changes to our assumed values can be readily incorporated in 
the framework laid out in this paper. Other uncertainties arise 
from our method of simulation; in particular, from our assumption of a static 
density profile for the protogalactic gas. We hope to address these issues
in future work. 

\section*{Acknowledgments}
The authors would like to acknowledge detailed comments by the anonymous 
referee which helped to significantly improve this paper. SCOG would also 
like to acknowledge useful discussions on various aspects of this work with 
Tom Abel, Omar Almaini, Rennan Barkana, Marie Machacek and Si Peng Oh. 
Financial support for much of this work was provided by a PPARC studentship. 
Additional support was provided by NSF grant AST99-85392.

%%\bibliographystyle{scognatbib}
%%\bibliography{bibfile}

\end{document}